\documentclass[%
 aip,
 jmp,%
 amsmath,amssymb,
 reprint,
 twocolumn,%
 floatfix,
]{revtex4}
\usepackage{graphicx}% Include figure files
\usepackage{dcolumn}% Align table columns on decimal point
\usepackage{bm}% bold math
\usepackage{subfig}
\usepackage{hyperref}

\begin{document}

\title[CONQUEST: Large scale DFT]{Large scale and linear scaling DFT
  with the CONQUEST code}

\author{Ayako Nakata}
\affiliation{International Centre for Materials Nanoarchitectonics
   (MANA), National Institute for Materials Science (NIMS), 1-1
   Namiki, Tsukuba, Ibaraki 305-0044, Japan}

\author{Jack S. Baker}
\affiliation{London Centre for Nanotechnology, University College
  London, 17-19 Gordon St, London, WC1H 0AH, UK}
\affiliation{Department of Physics \& Astronomy, University College
  London, Gower St, London, WC1E 6BT, UK}

\author{Shereif Y. Mujahed}
\affiliation{London Centre for Nanotechnology, University College
  London, 17-19 Gordon St, London, WC1H 0AH, UK}
\affiliation{Department of Physics \& Astronomy, University College
  London, Gower St, London, WC1E 6BT, UK}

\author{Jack T. L. Poulton}
\affiliation{London Centre for Nanotechnology, University College
  London, 17-19 Gordon St, London, WC1H 0AH, UK}
\affiliation{Department of Physics \& Astronomy, University College
  London, Gower St, London, WC1E 6BT, UK}

\author{Sergiu Arapan}
\altaffiliation{Present address: IT4Innovations, V\v{S}B - Technical University of
  Ostrava, 17. listopadu 2172/15, 70800 Ostrava-Poruba, Czech
  Republic}
\affiliation{International Centre for Materials Nanoarchitectonics
   (MANA), National Institute for Materials Science (NIMS), 1-1
   Namiki, Tsukuba, Ibaraki 305-0044, Japan}

\author{Jianbo Lin}
\affiliation{International Centre for Materials Nanoarchitectonics
   (MANA), National Institute for Materials Science (NIMS), 1-1
   Namiki, Tsukuba, Ibaraki 305-0044, Japan}

 \author{Zamaan Raza}
 \altaffiliation{Present address: Jülich Centre for Neutron Science (JCNS)
at Heinz Maier-Leibnitz Zentrum (MLZ), Lichtenbergstraße  1, 85748
Garching, Germany}
\affiliation{International Centre for Materials Nanoarchitectonics
   (MANA), National Institute for Materials Science (NIMS), 1-1
   Namiki, Tsukuba, Ibaraki 305-0044, Japan}

 \author{Sushma Yadav}
\affiliation{International Centre for Materials Nanoarchitectonics
   (MANA), National Institute for Materials Science (NIMS), 1-1
   Namiki, Tsukuba, Ibaraki 305-0044, Japan}
 
 \author{Lionel Truflandier}
\affiliation{Institut des Sciences Mol\'{e}culaires, Universit\'{e}
  Bordeaux, 351 Cours de la Lib\'{e}ration, 33405 Talence, France}

\author{Tsuyoshi Miyazaki}
\email{MIYAZAKI.Tsuyoshi@nims.go.jp}
\affiliation{International Centre for Materials Nanoarchitectonics
   (MANA), National Institute for Materials Science (NIMS), 1-1
   Namiki, Tsukuba, Ibaraki 305-0044, Japan}

\author{David R. Bowler}
\email{david.bowler@ucl.ac.uk}
\affiliation{International Centre for Materials Nanoarchitectonics
   (MANA), National Institute for Materials Science (NIMS), 1-1
   Namiki, Tsukuba, Ibaraki 305-0044, Japan}
\affiliation{London Centre for Nanotechnology, University College
  London, 17-19 Gordon St, London, WC1H 0AH, UK}
\affiliation{Department of Physics \& Astronomy, University College
  London, Gower St, London, WC1E 6BT, UK}

\date{\today}
\begin{abstract}
We survey the underlying theory behind the large-scale and linear
scaling DFT code, \textsc{Conquest}, which shows excellent parallel
scaling and can be applied to thousands of atoms with diagonalisation,
and millions of atoms with linear scaling.  We give details of the
representation of the density matrix and the approach to finding the
electronic ground state, and discuss the implementation of molecular
dynamics with linear scaling.  We give an overview of the performance
of the code, focussing in particular on the parallel scaling, 
and provide examples of recent developments and applications.
\end{abstract}

\keywords{DFT, linear scaling}

\maketitle

\section{\label{sec:intro}Introduction}
As computing power has increased and methods have become more
sophisticated, so computational modelling of materials systems has
taken its place alongside experiment and theory in science.
Electronic structure methods give insight into bonding and electronic
properties of systems, and density functional theory (DFT) is now the
\emph{de facto} method used in fields as diverse as physics,
chemistry, earth sciences, materials science and
biochemistry\cite{Jones:2015vt}.

However, almost all DFT calculations are focussed on a relatively
small system size, of a few hundred atoms (while feasible,
calculations involving more than a thousand atoms are still considered
expensive).  This size limitation mainly comes from the cost and scaling of
standard DFT implementations (asymptotically, the computer time required scales
with the cube of the number of atoms in the system, while the memory
scales with the square of the number of atoms).

The use of increasingly large numbers of computational cores is one route
to larger scale DFT calculations.  Indeed, high-performance
computing centres have recently scaled to tens of thousands and hundreds of
thousands of cores, and exascale computing is scheduled to arrive in the
next few years.  However, efficient use of systems of this size
requires care to ensure that the parallel scaling of any given code
remains efficient as  the number of cores becomes large.  
A real-space formulation, and the use of local basis functions to
represent the Kohn-Sham (KS) eigenstates, can help with the parallel
efficiency of the code, and enables calculations on larger systems,
often with several thousand
atoms\cite{Soler2002,Bowler:2002pt,Skylaris:2005rc,Rayson:2009vo,Bowler2010,VandeVondele:2012zl,Mohr:2015kb,Ozaki:2018vs,RomeroMuiz2018}.
Moreover, there is no inherent reason why DFT 
implementations should scale with the cube of the system
size\cite{kohn_density_1996}; in 
fact, with local basis functions, DFT can be formulated
in a linear scaling approach\cite{Goedecker:1999pv,Bowler:2012zt}.

The use of a few hundred atoms in a calculation potentially introduces
many approximations, and in some cases errors.  There are systems
where larger scale calculations are necessary to model the properties
correctly.  Features with dimensions of nanometres or more, or
features with strain fields over nanometres, will be seriously
restricted by a small calculation as the strain will not be fully
relaxed.  Dilute alloys can change their 
properties if the concentration is incorrect (for instance the
metal-insulator transition in doped silicon occurs at around $2\times
10^{19}$ dopants per cubic centimetre, or one dopant per 50,000
silicon atoms).  Calculations on biomolecules often use a small
quantum mechanical (QM) region embedded into a larger forcefield (or
molecular mechanics, MM)
region, and it has been shown\cite{Cole:2016im} that the
size and choice of the QM region affects the results strongly, with
sizes of at least 500 atoms being required.  While there are accurate
forcefields available, and new approaches to the fitting of
forcefields are being developed, simulations requiring electronic
structure or bond making and breaking must use DFT or a related approach.

\textsc{Conquest} is a large-scale and linear scaling DFT code which
is designed to scale efficiently in parallel, and to be applicable to systems with
thousands of atoms with full diagonalisation for the ground state,
and to systems with millions of atoms with a linear scaling solution
of the ground state.  It has recently been released under an
open-source MIT licence\cite{Conquest:2020aa}, and in this paper we describe the 
implementation of the code and various recent applications.  We first describe
the approaches to representing the density matrix, and finding the
electronic ground state, and then consider how to eigenvectors can be
calculated for systems large enough to require linear scaling.  We
also discuss the calculation of exact exchange with linear scaling.
We then turn to the movement of atoms, considering in particular the
calulation of stress, and the implementation of molecular dynamics,
both with reference to linear scaling.  After presenting the
performance of \textsc{Conquest} for various systems, we then
illustrate several applications with many thousands of atoms, before
concluding. 

\section{Methods: Electronic structure}
\label{sec:methods-elec-struc}

\subsection{Pseudopotentials and pseudo-atomic orbitals}
\label{sec:pseudopotentials}

The default pseudopotential format used in \textsc{Conquest} is the
optimised norm-conserving Vanderbilt pseudopotential (ONCVPSP)
developed by Hamann\cite{Hamann:2013kg} from Vanderbilt's
approach\cite{Vanderbilt:1985gy}.  This approach has been used to
generate two libraries covering most of the periodic table:
PseudoDojo\cite{Setten:2018xl} and SG15\cite{Schlipf:2015bw}.
Both of these libraries give very good values using the Delta
comparison to all-electron results\cite{Lejaeghere:2016ls}, with
PseudoDojo showing an accuracy comparable to the best PAW (projector
augmented waves) results
(when using a fully-converged plane wave basis set).

This approach to pseudopotential generation is focussed on 
accuracy, and as a consequence includes semi-core states for many
elements beyond the third row of the periodic table, as well as
partial core corrections\cite{Louie:1982ax} in many cases.  The
spacing of the real-space integration grid required for these
pseudopotentials may be finer than other approaches would give.
\textsc{Conquest} is also compatible with pseudopotentials in the
Troullier-Martins form as generated by SIESTA, which are less
stringent, and thus less expensive.

As is common for local orbital approaches, we use a neutral atom
approach\cite{Sankey:1989gf}, where the local part of the pseudopotential and a (neutral)
atomic density cancel each other out at large distances from the atom;
among other things, this removes the need for an Ewald sum for the
ion-ion energies.  However, it has been shown\cite{Ozaki:2005dp} that
the resulting potential can be deep, and require a fine integration
grid for convergence.  To alleviate this problem, we have implemented
the neutral atom projector 
approach\cite{Ozaki:2005dp}, and will report detailed testing in a
future publication.

\textsc{Conquest} can use pseudo-atomic orbitals (PAOs) as basis
functions to represent the support functions (discussed
  in Sec.~\ref{sec:repr-dens-matr} below), while the valence
orbitals also serve to generate the atomic density.
We generate the PAOs\cite{Bowler:2019fv} by solving the
Schr\"{o}dinger equation for an isolated, confined atom with a
pseudopotential.  The basis sets are formed from valence orbitals
(with a given number of functions, or zetas, for each angular momentum, each
with its own confinement) and polarisation functions (typically with
$l_{v}+1$ where $l_{v}$ is the angular momentum of the highest
occupied state, and a given, different number of functions to the
valence states).

The confinement is equivalent to a hard wall, and can be applied
either as a radial cutoff, or an energy shift for the orbital, which
generates a different radius for each orbital.  The
default basis sets in \textsc{Conquest} are generated either with the
same radial cutoffs for all shells, or the same energy shifts.  For
the energy shifts, we use a tight confinement (a shift of 2eV) and a
loose confinement (a shift of 0.02 eV) to generate two zeta functions,
with a third generated when needed using the average radius of the
first two.  The uniform radial confinement is found as the average of the
radii for all valence functions with the large or small energy confinement,
with the third radius again found as an average.  Semi-core states
only use one function with a loose confinement (in this case the
orbital is strongly confined so that even a very small energy shift
gives good results).

We have tested our default basis sets (single zeta with polarisation,
SZP, double zeta with polarisation, DZP, and triple zeta with triple
polarisation, TZTP) against converged plane wave calculations using
the same pseudopotentials\cite{Bowler:2019fv}.  We used a wide variety
of systems: elemental semiconductors (C, Si, Ge); oxides (SiO$_{2}$ in
both $\alpha$-quartz and stishovite phases, MgO, SrTiO$_{3}$, PbTiO$_{3}$ and
MgSiO$_{3}$); non-magnetic bcc Fe; and weakly bonded systems (ice XI and BN
sheets).  In all cases, we showed that the TZTP basis sets reproduced
the converged plane wave results with excellent accuracy: better than
1\% in bulk modulus and 0.1\% of the lattice constant, for a variety
of bulk systems; full details are found in Ref.~\onlinecite{Bowler:2019fv}.

\subsection{Representing the density matrix}
\label{sec:repr-dens-matr}

\textsc{Conquest} uses the density matrix as the fundamental
description of the system being modelling (in contrast to the
wavefunctions as is common in many DFT codes).  The density matrix is
represented in terms of support
functions\cite{Hernandez:1996bf,Hernandez:1995hc}, 
$\phi_{i\alpha}(\mathbf{r})$, where $i$ indexes the
  atom and $\alpha$ the function on the atom, and can be written:

  \begin{equation}
    \label{eq:1}
    \rho(\mathbf{r},\mathbf{r}^{\prime}) = \sum_{i\alpha, j\beta}
    \phi_{i\alpha}(\mathbf{r}) K_{i\alpha, j\beta} \phi_{j\beta}(\mathbf{r}^{\prime})
  \end{equation}
(Note that the density matrix can easily be written in terms of the
wavefunctions, when these are available, or found by linear scaling
optimisation, as described in Section~\ref{sec:solving-ground-state}.
Note also that, while we assume a non-spin-polarised
  calculation here, the extension to spin polarisation is
  straight-forward and has been implemented in \textsc{Conquest}.)

The support functions are local functions which move with the
atoms, and are strictly localised within a sphere.  They are used to
form the Hamiltonian and overlap matrices as well as to represent the density matrix:
  \begin{equation}
    \label{eq:2}
    H_{i\alpha,j\beta} = \int \mathrm{d}\mathbf{r}
    \phi_{i\alpha}(\mathbf{r}) \hat{H}
    \phi_{j\beta}(\mathbf{r}) 
  \end{equation}

The support functions themselves are defined either as primitive PAOs
(in a one-to-one mapping), or are represented in terms of one of two
basis sets: the PAOs; or blip
functions\cite{Hernandez:1997ay}.  We write:
\begin{equation}
  \label{eq:5}
  \phi_{i\alpha}(\mathbf{r}) = \sum_{s} b_{i\alpha s} \theta_{s}(\mathbf{r})
\end{equation}
where the basis functions $\theta_{s}(\mathbf{r})$ (either
pseudo-atomic orbitals or cubic B-splines) are discussed
further below.

\subsubsection{Multi-site support functions}
\label{sec:multi-site-support}

Since the primitive PAOs are localized around the atoms, we can use
them as support functions without any modifications, and a large PAO
basis set gives high accuracy, as shown in
Sec.~\ref{sec:pseudopotentials}.  However, the computational cost of
calculations scales with the cube of the number of support functions.
For large-scale calculations, we need to reduce the number of support
functions as much as possible.  We note that there is a strong link
between the basis set chosen and the number of support functions that
can be used\cite{Torralba:2008wm}.

Multi-site support functions (MSSF)\cite{Nakata:2014ev,Nakata:2015ld}
offer one way to reduce the number of support functions. The MSSF are
constructed as linear combinations of the PAOs not only on the
target atom, but also on its neighbouring atoms, defined by a cutoff
radius $r_{\mathrm{MS}}$, 

\begin{equation}
  \label{eq:MS1}
  \phi_{i \alpha}(\mathbf{r}) = \sum_{k}^{i, neighbours} \sum_{\mu \in k} C_{i \alpha, k \mu} \chi_{k \mu} (\mathbf{r}),
\end{equation}
where $\chi_{k\mu}$ is a PAO on atom $k$,
$\alpha$ is the index of the support functions of atom $i$,
$\mu$ is the index of the PAOs of its neighbouring atoms $k$
(including $i$ itself), and $C_{i\alpha,k\mu}$ are the linear
combination coefficients. Since the MSSF are no longer atomic
orbitals but local molecular orbital (MO)-like functions, the number
of MSSF can be equal to that of a minimal basis.  

In \textsc{Conquest}, two methods have been implemented to determine the
linear-combination coefficients of the MSSF. One of the methods is
the local-filter-diagonalisation (LFD) method  proposed by Rayson and
Briddon \cite{Rayson:2009vo,Rayson:2010fk,Nakata:2014ev}, and the
other is numerical optimisation\cite{Nakata:2015ld}, which will
be explained in Sec.~\ref{sec:solving-ground-state}.
In the LFD method, as shown in Eq.~\ref{eq:MS2}, the
  coefficients $\mathbf{C}$ for each atom are
found by diagonalising a subspace Hamiltonian matrix defined by a cluster
of radius $r_{\mathrm{LFD}}$ centred on the atom, constructed with PAOs; the coefficients,
$\mathbf{C}_{\mathrm{sub}}$, for the resulting eigenstates,
or local molecular orbitals,
are projected onto trial support 
functions, $\mathbf{t}$, localised on the target atom:
\begin{equation}
  \label{eq:MS2}
  \mathbf{C} = \mathbf{C}_{\mathrm{sub}} f(\epsilon_{\mathrm{sub}}) \mathbf{C}^\mathsf{T}_{\mathrm{sub}} \mathbf{S}_{\mathrm{sub}} \mathbf{t},
\end{equation}
where $\mathbf{S}_{\mathrm{sub}}$ is the overlap matrix
and $f(\epsilon_{\mathrm{sub}})$ is a Fermi function with a local
Fermi level $\epsilon_{\mathrm{sub}}$, used to exclude
high energy unoccupied local molecular orbitals.
Since the MSSF will depend on the charge
density of the system, which will in turn depend on the MSSF, 
the linear-combination coefficients need to be
determined self-consistently \cite{Nakata:2014ev}. $r_{\mathrm{MS}}$
should be equal to or smaller than the subspace region in the LFD method
$r_{\mathrm{LFD}}$. 

\begin{figure}
\centering
\includegraphics{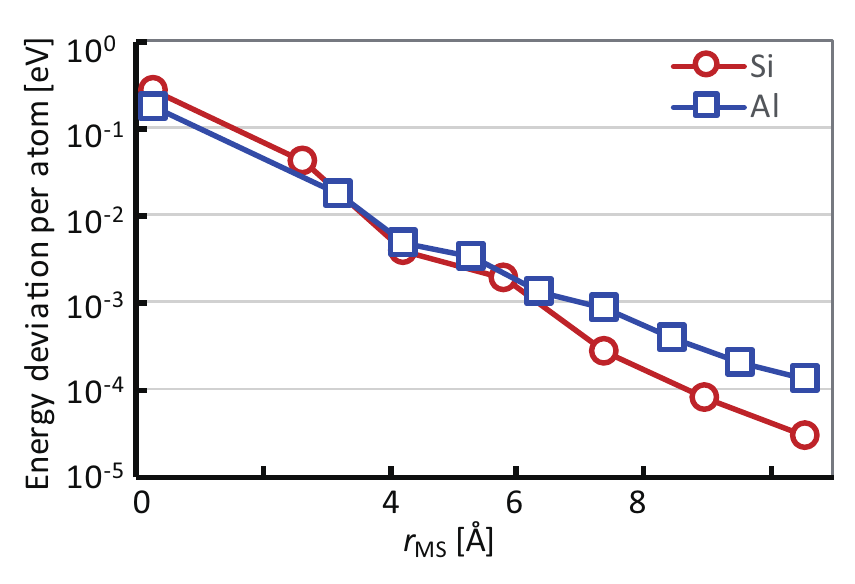}
\caption{Difference of total energy per atom [eV] with MSSF
  from the full primitive PAO result, with respect to the multi-site range
  $r_{\mathrm{MS}}$. The local diagonalization range
  $r_{\mathrm{LFD}}$ was set to be equal to $r_{\mathrm{MS}}$. The
  circles and squares correspond to bulk Si and
  Al. (Data taken from J. Chem. Theory Comput. 10, 4813 (2014) with
  permission.  Copyright (2014) American Chemical Society.)} 
\label{fig:MS1}
\end{figure}

The accuracy of the MSSF depends on $r_{\mathrm{MS}}$. In
Fig.~\ref{fig:MS1}, we see that the deviation from the full,
unrestricted, primitive PAO result decreases exponentially not only in
a gapped system (bulk Si) but 
also in a metallic system (bulk Al). The number of MSSF per atom is
four, while that of the TZP PAOs is 17 in both Si and Al, giving a
four-fold reduction in number and a significant speed-up. An example
of the computational time with the MSSF is demonstrated in
Sec.~\ref{subsec:performMSSF}. 

\subsubsection{On-site support functions}
\label{sec:site-supp-funct}

When using a linear scaling solver, as described in
Sec.~\ref{sec:solving-ground-state}, we require a sparse approximation
to the inverse overlap matrix to act as a metric\cite{White:1997hc}.
We have found that 
multiple zeta basis sets, and multi-site support functions, are not
compatible with our standard linear-scaling method for finding this
inverse overlap (Hotelling's method).  The reasons for this failure
are not yet clear, and are under investigation, but are most likely to
arise from the assumed sparsity pattern of the matrix, and the
starting value used\footnote{We note that the overlap matrix can be
  inverted without problem when using exact diagonalisation.}.
As a result, we have been limited in the basis
sets that can be used for linear scaling.  Blip functions, which will be
discussed in Sec.~\ref{sec:blip-functions}, offer an route to an accurate linear
scaling basis set; however, PAOs are often convenient and efficient,
and a restriction to SZ or SZP PAO basis sets is limiting.

We have found recently, however, that an adaptation of the MSSF
approach allows linear scaling solution for a sparse inverse overlap
matrix while retaining accurate basis sets: on-site support
functions (OSSF).  We restrict the PAOs forming the support functions for an
atom $i$ to its own PAOs; however, we must be careful to respect any
symmetry of the atomic lattice, so that the space spanned by the
support functions of the atoms decomposes into complete irreducible
representations of the symmetry group\cite{Torralba:2008wm}.  The
simplest way to ensure that this is respected is to increase the
number of support functions such that it encompases all angular
momenta of the PAOs (e.g. for PAOs including $l=0 \rightarrow 2$ we
would need 9 SFs, while for PAOs only including $l=0$ and $l=2$ we
would need 6 SFs).

This approach bears some similarity to the polarised atomic orbital
method\cite{Lee:1997tg,Berghold:2002fr} though that method imposes no
restrictions on the number of functions, and uses a different approach
to find the orbital coefficients.  In our approach we use the LFD
method described in Sec.~\ref{sec:multi-site-support}, using a trial vector which
is extended to include the polarisation orbitals.  We find that the
resulting support functions can be inverted efficiently
(interestingly, it is often more efficient than a simple SZP PAO basis
set).

When using OSSF with linear scaling, we are still investigating the
most efficient approach for finding the ground state; this involves
optimising the density matrix, the OSSF coefficients and the charge
density.  Introducing self-consistency between the OSSF coefficients and the charge density
is straightforward, but in a naive loop would add considerably to the
computational time.  Optimising the energy with respect to the OSSF
coefficients is also straightforward, but the most efficient approach
(i.e. when to update which parts of the optimisation) 
requires further research.

\begin{figure}[h]
  \centering
  \includegraphics[width=0.95\linewidth]{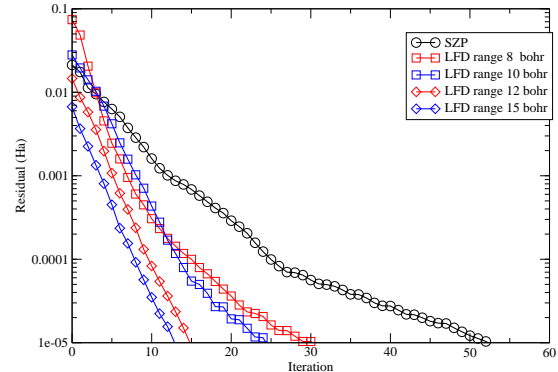}
  \caption{Convergence of linear scaling density matrix optimisation
    for different basis sets: SZP; and OSSF with different LFD
    ranges.  The system considered is an eight atom bulk silicon cell,
    slightly disturbed from the perfect crystal structure.}
  \label{fig:OSSF}
\end{figure}

The basis sets found using OSSF are significantly better than the
simple SZP PAO basis set, as shown in Fig.~\ref{fig:OSSF}.  Here we
see that, for a slightly disturbed eight atom bulk silicon cell,  as
the LFD range is progressively increased, so the rate of 
convergence of the density matrix optimisation improves.  The quality
of the resulting approximate inverse overlap matrix is also improved, and the
energy and forces on the atoms are significantly better with
the OSSF basis sets.  We show results 
for different basis sets in Table~\ref{tab:OSSFperformance}:
primitive PAOs (SZP, DZP and TZTP); MSSF for different ranges; and
OSSF.  The MSSF and OSSF calculations do not update the SF
coefficients after finding self-consistency; for MSSF, the LFD range
is set to 15 bohr throughout.  Note that the energy is not variational
with respect to the radius in this simple process, though with
optimisation (as described in Sec.~\ref{sec:optim-supp-funct}) it will
be variational.

\begin{table*}[h]
%  \centering
  \begin{tabular}{cccc}
    Basis             & Energy (Ha) & Force (Ha/bohr) & Time relative to TZTP\\
    \hline
    SZP               & -33.714     & -0.00170        & 0.12\\
    DZP               & -33.819     & -0.00151        & 0.23\\
    TZTP              & -33.838     & -0.00145        & 1.00\\
    \hline
    MSSF  5 bohr      & -33.800     & -0.00154        & 0.68\\
    MSSF  8 bohr      & -33.821     & -0.00143        & 0.73\\
    MSSF 10 bohr      & -33.818     & -0.00143        & 0.73\\
    MSSF 12 bohr      & -33.828     & -0.00144        & 0.81\\
    \hline
    OSSF  8 bohr      & -33.625     & -0.00162        & 0.63\\
    OSSF 10 bohr      & -33.813     & -0.00142        & 0.63\\
    OSSF 12 bohr      & -33.812     & -0.00142        & 0.60\\
    OSSF 15 bohr      & -33.820     & -0.00142        & 0.86\\
    \hline
    OSSF $\mathcal{O}(N)$  8 bohr & -33.605     & -0.00174        & 4.20\\
    OSSF $\mathcal{O}(N)$ 10 bohr & -33.784     & -0.00154        & 3.68\\
    OSSF $\mathcal{O}(N)$ 12 bohr & -33.782     & -0.00154        & 2.87\\
    OSSF $\mathcal{O}(N)$ 15 bohr & -33.792     & -0.00154        & 2.47\\
  \end{tabular}
  \caption{Comparison of different basis sets for an eight atom bulk
    silicon cell,
  slightly disturbed from the perfect crystal structure.  Primitive
  basis sets have 9, 13 and 27 support functions, respectively; MSSF have 4 support
  functions and a LFD range of 15 bohr; OSSF have 9 support
  functions.  MSSF and OSSF are not updated after the initial
  calculation of the coefficients.  The range on the density matrix
  for the $\mathcal{O}(N)$ calculations was 16 bohr.}
  \label{tab:OSSFperformance}
\end{table*}

As seen in Fig.~\ref{fig:OSSF}, it is evident that the OSSF give a
significant improvement to the performance of the linear scaling
solver, and from Table~\ref{tab:OSSFperformance} we can see that they
are comparable to the MSSF for the accuracy of forces and 
the timing.  Most notably, we see that with the OSSF found with the
LFD radius set to 15 bohr, the linear scaling solver is only 2 times
slower than diagonalisation with the full TZTP basis set, and 8
times slower than the DZP basis set.  This performance difference is
expected for such a small system where linear scaling solvers are less
efficient than diagonalisation and related solvers.  However, it suggests that the choice of
basis functions is important in implementing linear scaling.  We note that the quality of both 
OSSF and MSSF basis sets would be improved by optimisation of the
coefficients, as described in Sec.~\ref{sec:optim-supp-funct}, but
even with these simple approximations, good performance is achieved.

\subsubsection{Blip functions}
\label{sec:blip-functions}

While PAOs are a convenient basis set, they do not permit systematic
convergence of the energy with respect to the basis: while adding
extra basis functions will increase the size of the variational space
and lower the energy,
there are two parameters which offer different degrees of 
freedom (maximum angular momentum, and number of radial functions, or
zetas, per angular momentum channel), and there are no guarantees of
how adding to each parameter will change the energy.

The blip functions\cite{Hernandez:1997ay}, which are piecewise
continuous cubic splines defined on a cubic grid that moves with the
atoms, are a basis set that can be systematically converged.  The blip
grid spacing corresponds directly to a plane wave cutoff, allowing the
basis set to be improved systematically (of course, the support functions
are confined within a radius, but it has been shown that the total
energy converges variationally and rapidly with this
radius\cite{Hernandez:1997ay,Bowler:2000xd}).  The most efficient
procedure for initialisation and optimisation of the blip coefficients
along with the charge density, and for linear scaling approaches to
finding the density matrix, is the subject of on-going research.

\subsection{Solving for the ground state}
\label{sec:solving-ground-state}

The ground state electronic structure in \textsc{Conquest} is defined
by three related quantities: the support functions; the density
matrix; and self-consistency between the charge density and the
Kohn-Sham potential.  The self-consistency procedure is a standard
part of DFT and related codes\cite{Woods:2019yo}, and we implement the
Pulay approach\cite{Pulay:1980zi} which works efficiently.

While the overall search for the ground state could be considered as
an optimisation in a space formed by both the support function
coefficients and the density matrix elements, it is easier to consider
how the density matrix is found for a given set of support functions,
and then to discuss methods for optimising the energy with respect to
the support function coefficients.  We consider first the two
approaches to solving for the density matrix: exact
diagonalisation\footnote{As is common in the field, we
    use the term ``exact diagonalisation'' throughout the paper to
    indicate that the eigenstates and density matrix are found without
    search or approximation.}, which scales cubically but makes no
approximations; and linear scaling, which imposes a range on the
density matrix.

\subsubsection{Density matrix: exact diagonalisation}
\label{sec:dens-matr-exact}

We perform diagonalisation of the Hamiltonian using
ScaLAPACK\cite{Blackford:1997br}, and are also investigating the use
of ELPA\cite{Marek:2014aa} (which uses the same interface, and may
scale better to large numbers of processes).

Since we apply periodic boundary conditions to our simulation cell,
the Brillouin zone must be sampled appropriately; we have implemented
the Monkhorst-Pack\cite{Monkhorst:1976aa} sampling method as a default
approach to Brillouin zone sampling, but any arbitrary set of k-points
can be used.  (At present we do not account for the symmetry of the
simulation cell beyond time-reversal symmetry, as the code is designed
for large-scale simulations which are unlikely to show significant
symmetries.)

The Kohn-Sham eigenstates are represented in terms of the support
functions, with the density matrix found as: 
\begin{eqnarray}
  \label{eq:3}
  \mid \psi_{n\mathbf{k}}\rangle &=& \sum_{i\alpha} c^{n\mathbf{k}}_{i\alpha} \mid
                                     \phi_{i\alpha}\rangle\\
  K_{i\alpha, j\beta} &=& \sum_{n\mathbf{k}} f_{n\mathbf{k}} w_{\mathbf{k}}
                          c^{n\mathbf{k}}_{i\alpha} \left(c^{n\mathbf{k}}_{j\beta}\right)^{\star}
\end{eqnarray}
where the weight of each k-point is given as $w_{\mathbf{k}}$ and the
occupancy of the eigenstate is $f_{n\mathbf{k}}$ (which is found
using a simple Fermi-Dirac distribution, or the
Methfessel-Paxton\cite{Methfessel:1989cx} approach).  The
diagonalisation at each k-point can be assigned to a sub-group of
processes, enabling a calculation using many k-points to be sped up
significantly. 

\subsubsection{Density matrix: linear scaling}
\label{sec:dens-matr-line}

To achieve linear scaling in computational time with the system size,
we restrict the range of the density matrix (the range
  of a matrix $A_{i\alpha, j\beta} = 0$ is defined in terms of the
  distance between atoms $i$ and $j$,
  $R_{ij} = \mid \mathbf{R}_{i} - \mathbf{R}_{j}\mid$, and is
  restricted by setting matrix elements to zero when $R_{ij}$ is greater
  than a cutoff distance $R_{c}$), and optimise the band energy,
$E_{\mathrm{band}} = 2\mathrm{Tr}[KH]$, with respect to the density
matrix elements.  When this approach 
is coupled with strictly local basis functions, all matrices are
sparse, and all matrix operations scale linearly with system size.

During the optimisation, we must constrain the electron number (a
relatively straightforward task\cite{Hernandez:1996bf}), and also the
idempotency of the density matrix (a much more complex task in a
variational context).  We follow the LNV (Li, Nunes and Vanderbilt) 
approach\cite{Li:1993lg,Nunes:1994pi} where we write the density
matrix $K$ in terms of an \emph{auxiliary} density matrix $L$, using the
McWeeny transform:
\begin{equation}
  \label{eq:4}
  K = 3LSL - 2LSLSL
\end{equation}

This drives the density matrix, $K$, towards idempotency (strictly it
is driven towards \emph{weak} idempotency, where the eigenvalues lie
between zero and one, but may not be exactly zero and one).  If, as
above, we write $E = \mathrm{Tr}[KH]$ then we can use the gradient $\partial
E/\partial L_{i\alpha, j\beta}$ to minimise the energy with respect to
the density matrix, and the density matrix $K$ will be driven towards
idempotency as the minimisation proceeds.  Here the
  range of $K$ is the same as the range of the Hamiltonian (which is
  naturally sparse); it is $L$ whose range is restricted, and this
  range sets the accuracy of the calculation.

The initial density matrix is generated from the Hamiltonian, using an
iterative procedure based on a generalisation of the McWeeny
transform\cite{Palser:1998fd,Bowler:1999if}.  We use an approximate,
sparse inverse overlap matrix as the metric for the optimisation,
found using the iterative Hotelling method.  As discussed in
Sec.~\ref{sec:repr-dens-matr}, this has certain consequences for the
basis sets that can be used, but both OSSF and blip basis sets show
promise for accurate, linear scaling calculations.

\subsubsection{Optimising support functions}
\label{sec:optim-supp-funct}

As mentioned in Sec.~\ref{sec:repr-dens-matr}, we can construct
support functions by taking linear combinations of PAOs (MSSFs or
OSSFs) or blips. The linear combination coefficients can be optimised
numerically by minimizing the DFT total energy with respect to the
coefficients\cite{Nakata:2015ld}. For MSSFs and OSSFs, the
coefficients obtained by the LFD method generally form good initial
values for the numerical optimisation.  The initial blip
coefficients are found as a best fit to PAOs.  In this section, we
demonstrate optimisation of SF coefficients for the MSSFs, though the
formalism is identical for the other approaches.  We note that these
optimisation processes are liable to ill-conditioning, which can be
mitigated\cite{Bowler:1998lo}.

\begin{figure*}
\centering
\includegraphics{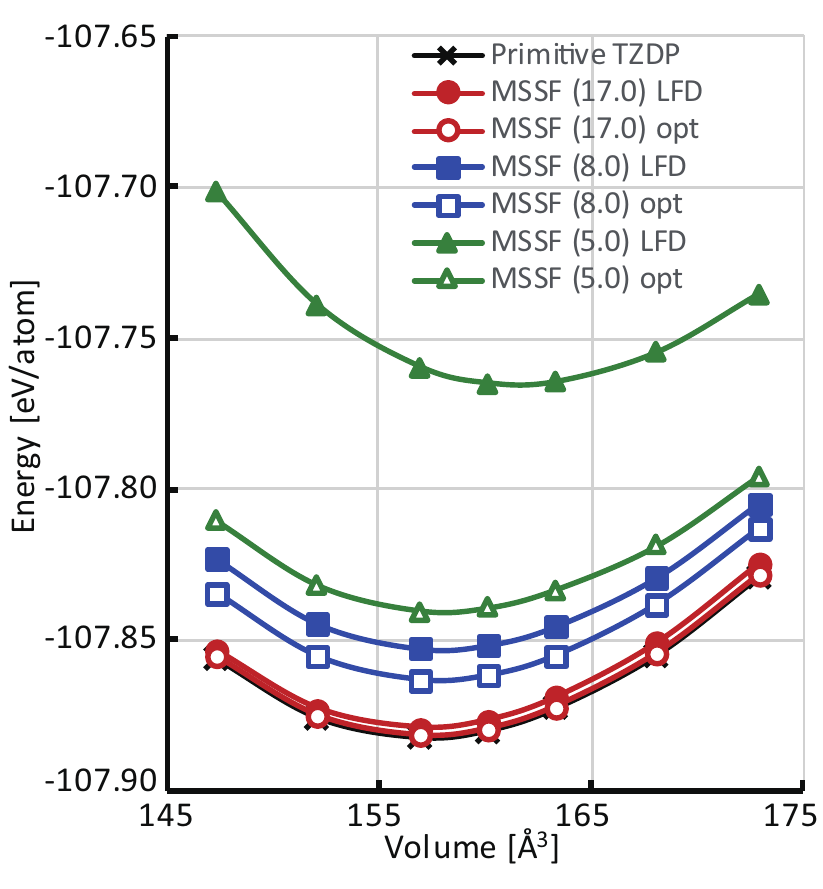}
\caption{Energy-volume curves of bulk Si, demonstrating the effect of
  optimising MSSF coefficients after the initial LFD process. Symbols
  correspond to the calculated energies by primitive TZDP PAOs and
  MSSF with multi-site ranges
  ($r_{\mathrm{MS}}$) 17.0, 8.0 and 5.0 bohr, using the LFD method
  (filled symbols) and numerical optimisation (open symbols). The
  local filter diagonalization range $r_{\mathrm{LFD}}$ was set to be
  equal to $r_{\mathrm{MS}}$. (Adapted from
  Ref.~\onlinecite{Nakata:2015ld} with permission from the PCCP Owner
  Societies.)}  
\label{fig:MS2}
\end{figure*}

Figure~\ref{fig:MS2} shows the energy-volume (E-V) curves of bulk
Si\cite{Nakata:2015ld} calculated with simple LFD (filled symbols) and
optimisation of the MSSF coefficients (open symbols).
The number of MSSFs per atom is four, while
that of the primitive TZDP (3s3p2d) PAOs is 22. Table~\ref{table:MS1}
summarises the lattice constant $a_0$ obtained by fitting the E-V
curves with the Birch-Murnaghan equation. The results are improved, i.e.,
becoming closer to the results of the primitive PAOs, by the numerical
optimisation in all cases. When $r_{\mathrm{MS}}$ is large, e.g. 17.0
bohr, since the MSSFs found with LFD give accurate results,
the change from numerical optimisation is small. When
$r_{\mathrm{MS}}$ is 8.0 bohr, the difference with and without the
numerical optimisation is significantly larger, but both LFD and the
numerical optimisation show 
reasonable accuracy. On the other hand, when $r_{\mathrm{MS}}$ is as
small as 5.0 bohr, the result with the LFD method is not accurate,
with a 1\%  deviation from the full TZDP result, but
we find significant improvement of the accuracy from numerical
optimisation, reducing the percentage deviation to 0.2\%.

\begin{table}
  \begin{tabular}{rcccc}
  \hline
  $r_{\mathrm{MS}}$ & \multicolumn{2}{c}{$a_0$} & \multicolumn{2}{c}{\%D} \\ 
       & LFD   & opt               & LFD & opt \\
    \hline
   5.0 & 5.447 & 5.406             & 1.0 & 0.2 \\
   8.0 & 5.403 & 5.400             & 0.2 & 0.1 \\
  17.0 & 5.393 & 5.395             & 0.0 & 0.0 \\
  TZDP & \multicolumn{2}{c}{5.395} & \multicolumn{2}{c}{---}\\
  \hline  \end{tabular}
  \caption{Lattice constants $a_0$ of bulk Si calculated with MSSF
    with ranges $r_{\mathrm{MS}}$ of 5.0, 8.0 and 17.0 bohr, and
    percent deviations (\%D) from $a_0$ calculated with the primitive
    TZDP PAOs.} 
  \label{table:MS1}
\end{table}

\subsection{Electronic structure for large systems}
\label{sec:electr-struct-large}

Linear scaling, or $\mathcal{O}(N)$, calculations which work with the density matrix implicitly
integrate over energy and produce only the sum of the occupied
eigenvalues and not any of the Kohn-Sham eigenstates of the
system. However, we often want to know individual eigenstates to
analyze the electronic structure of the system, though generally
within a relatively small energy range. These can be found efficiently
from the converged ground-state Hamiltonian by using the
Sakurai-Sugiura (SS) method\cite{Nakata:2017ss}. The SS method
\cite{Sakurai:2003ss,Sakurai:2013ss} is an efficient interior
eigenproblem solver for large sparse matrices using contour integrals
in the complex plane, which provides the eigenvalues and eigenvectors in a
finite eigenvalue range with high parallel
efficiency.  We use the SS method as it is much more
  scalable in parallel than other approaches such as shift-and-invert
  Lanczos\cite{Yamazaki2013}.) We first 
optimise the electronic Hamiltonian with the $\mathcal{O}(N)$ method in
\textsc{Conquest}, and then obtain the eigenstates of the Hamiltonian
in a finite energy window with a one-shot SS calculation. Here we
demonstrate the usefulness of the combination of the $\mathcal{O}(N)$ and SS
methods by showing two examples. 

\begin{figure*}
\centering
\includegraphics{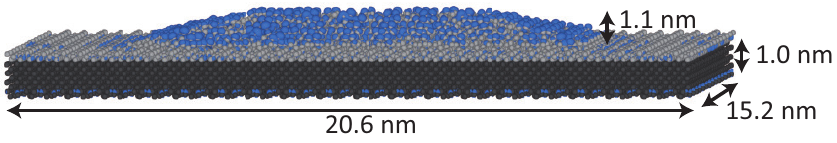}
\caption{Electronic density distributions (blue) of the Ge hut
  clusters (light gray) on the Si(001) (dark gray) (totally 23,737
  atoms) around the Fermi level. (Reprinted with permission from
  J. Chem. Theory Comput. 13, 4146 (2017). Copyright (2017) American Chemical
    Society.)}
\label{fig:SS1}
\end{figure*}

The first example is the energy-specific electron-density distribution
in a hut-shaped Ge cluster on Si(001) surface consisting of 23,737
atoms (the physical system is described in more detail in 
Sec.~\ref{subsec:AppSiGe}). Figure~\ref{fig:SS1} shows the electron
density distribution in the energy range [-0.01 eV: + 0.02 eV] around
the Fermi level,  obtained by calculating the Kohn-Sham
eigenvectors in this  range with the SS method. The calculation
for the eigenvalues and eigenvectors required 146 seconds using 64
nodes of the K supercomputer. We also calculated the
eigenstates in the same energy range for a larger Ge/Si(001) system,
consisting of 194,573 atoms, in 2,399 seconds using 6,400 nodes.
(Note that the times quoted are just for the SS
  eigensolutions, which are performed as post-processing calculations
  using output from \textsc{Conquest}).

\begin{figure*}
\centering
\includegraphics[width=\linewidth]{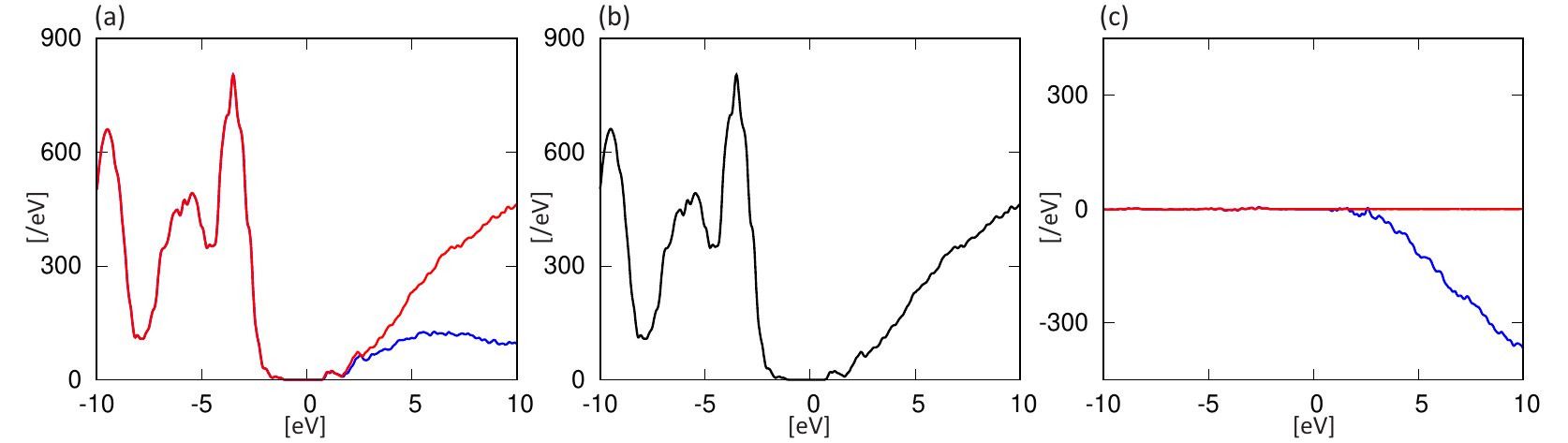}
\caption{Density of states of hydrated DNA obtained with (a)
  multi-site support functions (MSSF) with multi-site range 8.0 bohr
  (blue) and DZP PAOs using MSSF charge density (red) and (b) DZP PAOs
  using DZP SCF charge density (black). The difference of (a) MSSF
  (blue) and DZP with the MSSF density (red) from (b) DZP is also
  shown in (c). The eigenstates in (a) and (b) were obtained by the SS
  method. (Reprinted with permission from
  J. Chem. Theory Comput. 13, 4146 (2017). Copyright (2017) American Chemical
    Society.)} 
\label{fig:SS2}
\end{figure*}

The second example is the density of states (DOS) of a DNA system in
water, which consists of 3,439 atoms. The DOS calculated with MSSF
(see Sec.~\ref{sec:multi-site-support}) (4,774 functions) and 
primitive PAOs (27,883 functions), and their difference are shown in
Fig.~\ref{fig:SS2}. The DOS from the MSSF is very close to the full
primitive PAO DOS in the occupied states, and in the unoccupied states near
the Fermi level, while the DOS in the unoccupied states far from Fermi level
are quite different. This is because the MSSFs are determined by
optimising the occupied states with a small number of support functions,
and the accuracy of the MSSFs for unoccupied states often becomes
poor. To improve this poor description, we can use the SS
method. First, we optimise the electronic density of the target system
using MSSF, and we re-construct the electronic Hamiltonian using
the primitive PAOs with the optimised density. Then we use the SS
method to obtain the eigenstates. Thus, we can obtain the DOS even in
unoccupied states far from Fermi level quite accurately, as shown in
Fig.~\ref{fig:SS2}.

\subsection{Exact exchange}
\label{sec:exact-exchange}

Exact exchange (EXX) correction to the original Kohn-Sham formulation of DFT,
leading to the class of \textit{hybrid} exchange-correlation functionals,
has become very popular, since in the vast majority of cases, it improves
the overall reliability of the DFT predictions. Depending on the implementation 
--- mainly basis set and boundary conditions ---  orbital dependence
is introduced to the KS-DFT formalism \textit{via} the EXX energy standard expression:
\begin{widetext}
\begin{equation}
\label{eq:exx1}
% 	 E_{x}[\{\psi_n\}] 
 	 E_{x} 
 	 = -\frac{1}{4}\int\textrm{d}\mathbf{r}\textrm{d}\mathbf{r}^\prime \frac{\rho(\mathbf{r},\mathbf{r}^\prime) \rho(\mathbf{r}^\prime,\mathbf{r})}
 	 {\vert\mathbf{r} - \mathbf{r}^\prime\vert} = -\sum_{n,m}\int\textrm{d}\mathbf{r}\textrm{d}\mathbf{r}^\prime
 	 \frac{\psi^{*}_{m}(\mathbf{r})\psi^{*}_{n}(\mathbf{r}^\prime) \psi_{n}(\mathbf{r})\psi_{m}(\mathbf{r}^\prime)}
 	 {\vert\mathbf{r} - \mathbf{r}^\prime \vert},
\end{equation}
\end{widetext}
where $\{\psi_n\}$ is the set of $N$ occupied KS states, can bring the computational 
cost to a prohibitive level more rapidly than pure LDA/GGA DFT when increasing the
system size. Within the framework of \textsc{Conquest}, where the density matrix 
of Eq.~\ref{eq:exx1} are expanded onto a set of $M$ localized and real basis functions, 
the exchange energy reads: $E_{x} = -\textrm{Tr}\left\{KX\right\}$,
with the exchange matrix ($X$) elements given by:
\begin{widetext}
\begin{equation}
\label{eq:exx2}
  X_{i\alpha,j\beta} 
  =  \sum_{k\mu,l\nu}
  \int\textrm{d}\mathbf{r}\textrm{d}\mathbf{r}^\prime\frac{\rho_{i\alpha,k\mu}(\mathbf{r})K_{k\mu,l\nu}\rho_{l\nu,j\beta}(\mathbf{r}^\prime)}
  {\vert\mathbf{r} - \mathbf{r}^\prime\vert} 
  =  \sum_{k\mu,l\nu}
  \int\textrm{d}\mathbf{r}\textrm{d}\mathbf{r}^\prime\frac{\phi_{i\alpha}(\mathbf{r})\phi_{k\mu}(\mathbf{r})
  K_{k\mu,l\nu}\phi_{l\nu}(\mathbf{r}^\prime)\phi_{j\beta}(\mathbf{r}^\prime)}{\vert\mathbf{r} - \mathbf{r}^\prime\vert}.
\end{equation}
\end{widetext}
As a result, calculation of $X$ requires to evaluate at most $M^4$ 
electron repulsion integrals (ERIs) defined by the integrand of Eq.~\ref{eq:exx2}.
The first equality in the equation above outlines the fact that evaluating an ERI
is similar to computing the Hartree energy, with in place of the full electronic
density, localized pair densities $(\rho_{i\alpha,k\mu},\rho_{l\nu,j\beta})$ 
coupled by the density matrix elements $K_{k\mu,l\nu}$. Consequently, ERI 
calculation can be performed by solving a Poisson equation into a predefined 
local cell. Note that, contrary to the Hartree energy, solution of this equation 
should be free of periodic boundary conditions.

When dealing with a numerical basis set such as the PAOs, several options to compute 
the ERIs are available, with for instance the semi-analytic solution given by Toyoda 
and Ozaki\cite{toyoda_numerical_2009,toyoda_liberi_2010} combining fast-spherical 
Bessel transform for the radial integration and a more traditional analytic method for 
the spherical harmonic part. Another approach is based on the experience of Gaussian-type 
orbital (GTO) ERI solvers.\cite{Libint1} In that case, the PAO-ERIs are transformed into a 
set of contracted GTO-ERIs which are then calculated 
analytically.\cite{shang_implementation_2011,qin_honpas_2015} Instead, we use a route 
which circumvents the calculation of the ERIs and works for any smooth finite-range 
functions, which is particulary well suited for $\mathcal{O}(N)$ approaches based on the 
pseudopotential approximation. 
The key part is to perform the sum over the index $l\nu$ before solving for the Coulomb
potential of the pair densities; this simple re-ordering increases the efficiency of the 
procedure markedly. For this, we introduce the  contraction functions, $\Phi_{k\mu}(\mathbf{r}^\prime)$, as:
\begin{equation}
  \label{eq:exx3}
  \Phi_{k\mu}(\mathbf{r}^\prime) = \sum_{l\nu} K_{k\mu,l\nu}\phi_{l\nu}(\mathbf{r}^\prime)
\end{equation}
It should be noted that the domain over which these functions are defined requires some 
care.\cite{exx_arxiv} The sum over $l\nu$ need only include those support functions 
$\phi_{l\nu}$ overlapping with $\phi_{j\beta}$, as $\Phi_{k\mu}$ will be multiplied by this 
function. Contracted densities are then defined as
%:
\begin{equation}
  \label{eq:exx4}
  \bar{\rho}_{k\mu,j\beta}(\mathbf{r}^\prime) = \Phi_{k\mu}(\mathbf{r}^\prime)\phi_{j\beta}(\mathbf{r}^\prime),
\end{equation}
and the resulting Coulomb potential,
\begin{equation}
  \label{eq:exx5}
  \bar{v}_{k\mu,j\beta}(\mathbf{r}) = \int \textrm{d}\mathbf{r}^\prime \frac{\bar{\rho}_{k\mu,j\beta}(\mathbf{r}^\prime)}{\vert \mathbf{r} - \mathbf{r}^\prime \vert},
\end{equation}
is calculated by solving Poisson's equation. Once the potential has 
been found, a further contraction over $k\mu$ is performed to create,
\begin{equation}
  \label{eq:exx6}
  \Omega_{j\beta}(\mathbf{r}) = \sum_{k} \bar{v}_{k\mu,j\beta}(\mathbf{r}) \phi_{k\mu}(\mathbf{r}),
\end{equation}
where, again, the sum over support functions $k\mu$ need only include those 
functions which overlap with functions $i\alpha$.  The exchange matrix elements
are then calculated by numerical integration:
\begin{equation}
  \label{eq:exx7}
  X_{i\alpha,j\beta} = \int \textrm{d}\mathbf{r} \phi_{i\alpha}(\mathbf{r}) \Omega_{j\beta}(\mathbf{r}).
\end{equation}
The process ---from Eq.~\ref{eq:exx3} to ~(\ref{eq:exx7})--- by which the EXX is calculated
in \textsc{Conquest}, will be referred as to the contraction reduction integral (CRI).
The set of function $\Omega_{j\beta}$ is effectively defined by the density matrix range, 
and the need for $j\beta$ to overlap with atoms $l\nu$, which naturally control the number
of functions entering in the sums of Eqs.~(\ref{eq:exx4}) and ~(\ref{eq:exx6}).
Note that the calculation time can be reduced by imposing a range condition ($R_X$)
on the exchange matrix. This is related to the sparsity 
property\cite{kohn_density_1996} of $\rho(\mathbf{r},\mathbf{r}^\prime)$ and the truncation 
of all the operators involved in the Hamiltonian.\cite{Hernandez:1996bf} 
\begin{figure}[bht!]
  \centering
  \includegraphics[width=0.4\textwidth]{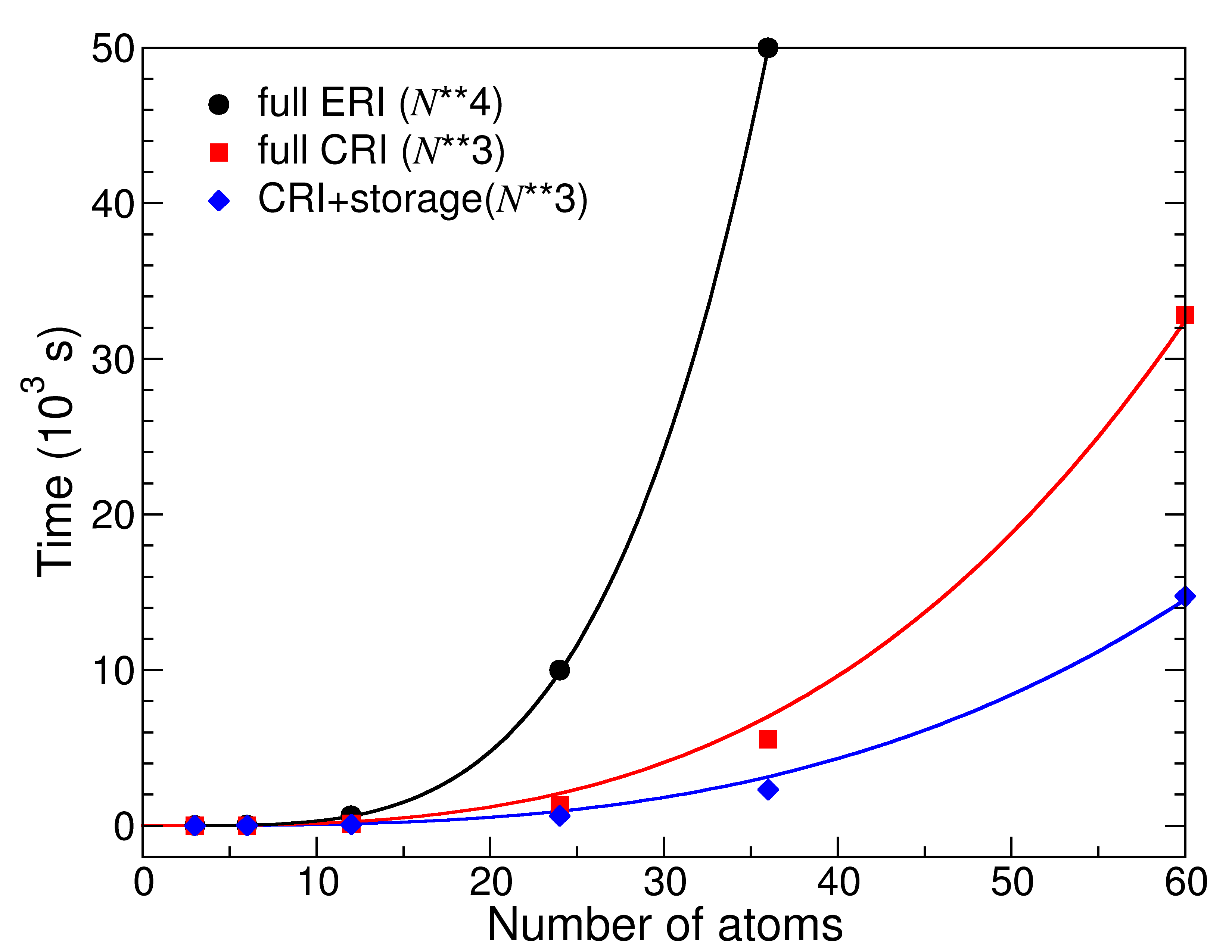}
  \caption{Comparison of CPU times necessary to compute EXX in 
	   isolated water clusters as a function of number atoms 
	   ($N$) using explicit ERI calculation and the CRI method. 
	   Ideal $N^4$ and $N^3$ scalings are given by plain lines.
	   \label{fig:scaling1}}
\end{figure}
\begin{figure}[bht!]
  \centering
  \includegraphics[width=0.4\textwidth]{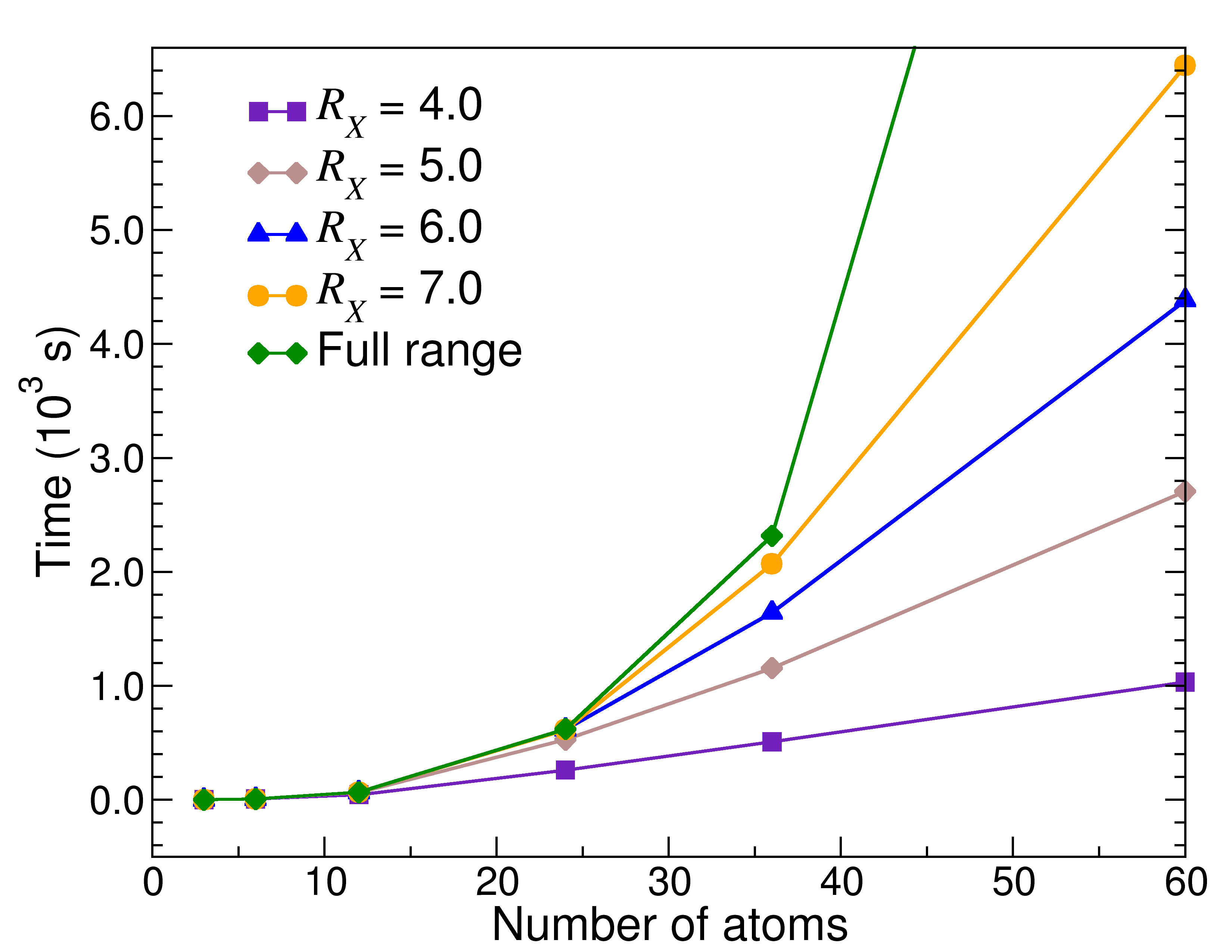}
  \caption{Variation of the CPU time with respect to the range $R_X$ (in a.u.) for
	   the calculation of EXX in isolated water clusters using the CRI 
	   method.\label{fig:scaling2}}
\end{figure}

Practical tests on the efficiency of this approach algorithm were 
carried out on a set of isolated water clusters (H$_2$O)$_n$ ($n\leq20$) with fused 
cubes structures.\cite{wales_global_1998}
Calculations of exchange energy were performed after the KS density matrix had been 
converged using the standard self-consistent-field (SCF) method. As a result, the 
timings presented below for exact exchange (EXX) energy can be compared to a 
single SCF cycle as found in hybrid-DFT calculation. For this demonstration, 
SZP PAO orbitals have been used for hydrogen and 
oxygen with cutoff radii of 4.7 and 3.8 au, respectively. 
We emphasize that the main conclusions of this work can be easily extended to more 
flexible basis sets, as long as the support functions are localized. 
The CPU times used for the computation of EXX are reported in Fig.~\ref{fig:scaling1}
as a function of the number of atoms using: (i) the explicit evaluation of the full set of 
ERI, (ii) the CRI approach, and (iii) the CRI approach with partial storage of the PAO 
on the grids. Comparing the formal scalings obtained for the CRI methods against 
the full ERI approach, it becomes clear that the former reduces the quartic scaling 
to cubic with respect to the size of the system.

At this point we should emphasise that exchange energy 
values obtained with the three schemes are fully identical, their accuracies being
only dependent on the Poisson solver used to evaluate the pair potential in Eq.~\ref{eq:exx5}.
Among the various numerical methods, one can choose to evaluate the Coulomb 
potential in reciprocal or real space. Whereas the former is the most appropriate for periodic
neutral systems ---when the positively charged nuclei compensate exactly 
the electronic charge density--- it becomes less reliable for isolated and/or 
charged systems.\cite{castro_solution_2003} Several schemes have been 
developed to tackle this problem,\cite{jarvis_supercell_1997,martyna_reciprocal_1999,rozzi_exact_2006,dabo_electrostatics_2008}
Alternatives based on the discrete variable representation (DVR) of~Eq.~\ref{eq:exx5} which 
avoids the direct resolution of the Poisson equation have been 
proposed.\cite{lee_efficient_2008} The density is generally expanded in 
a direct product of one-dimensional localized real-space basis 
functions\cite{watson_linear-scaling_2008,lee_efficient_2008,varga_lagrange_2004}
as for instance, interpolating scaling functions (ISF). After extended comparisons between the DVR-ISF 
developed by Genovese et al.\cite{genovese_efficient_2006,genovese_efficient_2007} and 
corrected FFT-based schemes,\cite{onida_ab_1995,blochl_electrostatic_1995,schultz_local_1999} 
we found that systematic convergence of the ERI is obtained with a better accuracy and at a lower 
cost using the real space Poisson solver.

\begin{figure}[tbh!]
  \centering
  \includegraphics[width=0.4\textwidth]{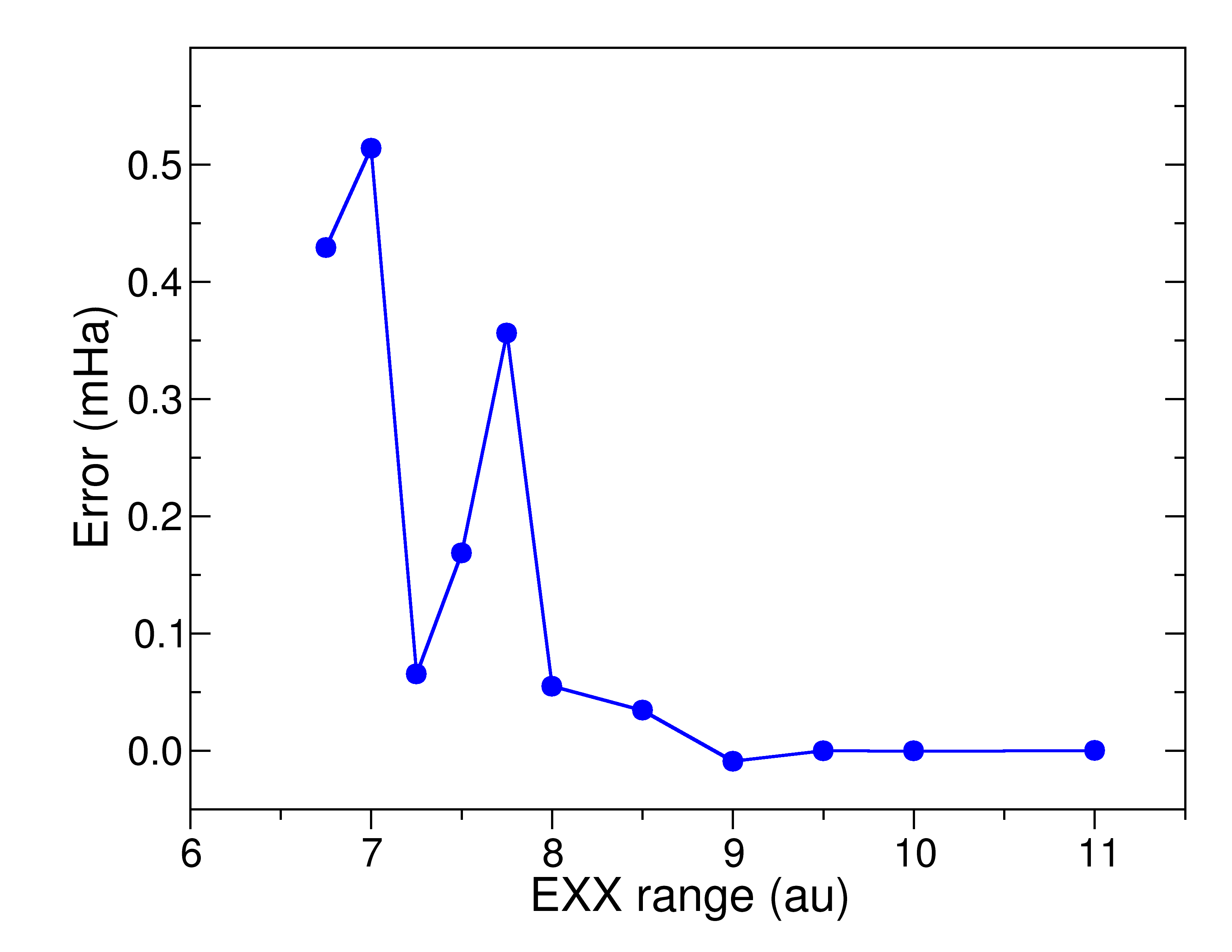}
  \caption{Convergence of the EXX energy with respect to the exchange 
	   range $R_X$ for the cluster (H$_2$O)$_{20}$. Error is given with respect 
	   to the exact calculation.\label{fig:scaling3}}
\end{figure}

As shown in Fig.~\ref{fig:scaling2}, if a finite range $R_{X}$ is introduced within the CRI 
algorithm, the CPU time can be significantly reduced, allowing linear
scaling to be achieved for 
clusters with more than 36 atoms (with $R_X=7.0$ au). Computational resources further 
decrease with shorter EXX ranges, along with faster onset of the linear-scaling regime. 
The EXX accuracy with respect to the range of the exchange matrix is shown in 
Fig.~\ref{fig:scaling3} for the cluster (H$_2$O)$_{20}$ presenting the ``boxkite'' 
structure. After somewhat erratic behavior at low values, it is found that an accuracy below 0.5 mHa 
is reached for $R_X\geq 8$. Even though the non-local nature of the EXX interaction may need some
special care when introducing a cutoff radius on $X$ elements, it is reasonable to believe 
that the CRI implementation, along with a judicious choice of convergence parameters,
is opening the way to exact exchange calculations on 100,000+ atoms with \textsc{Conquest}
for a fair efficiency/accuracy ratio.

\section{Methods: Moving atoms}
\label{sec:meth-moving-atoms}

\subsection{Forces and stresses\label{sec:forces-stresses}}

Forces have been available in \textsc{Conquest} for some time, as described
elsewhere\cite{Miyazaki:2004ee,Torralba:2009nr}, with the force being
the exact differential of the energy, including Pulay forces where
appropriate. 

Calculation of the stress tensor has recently been implemented within the
current release of \textsc{Conquest}.  The definition of
the stress tensor is standard\cite{Soler2002}: 

\begin{equation}\label{eq:gen-stress}
	\sigma_{\alpha\beta} = \frac{\partial E}{\partial
          \epsilon_{\alpha\beta}} = \frac{\partial E}{\partial r_{\alpha}}r_{\beta}
\end{equation}
where $\alpha$ and $\beta$ are Cartesian directions indices, and the
second equality holds for \emph{most} contributions to the stress.  In this case, the first term
is the force, so most contributions to the stress tensor can be
calculated at the same time the forces are calculated. There are a few
exceptions to this, however, but they are easily evaluated\cite{Soler2002,Knuth2015}. 

The original formulation of stress within DFT is traced back to the
pioneering work of Nielsen and Martin\cite{Nielsen1983,Nielsen1985},
where a formulation for the stress was expressed for the first time in
the framework of the local-density approximation (LDA) and later
derived in more detail\cite{Nielsen1985a,Nielsen1987}.  We have chosen to omit the
factor of $\frac{1}{\Omega}$ in eq. \ref{eq:gen-stress} since it
\textit{averages} the total stress over the macroscopic simulation
cell and in a case where the volume $\Omega$ is not well defined would
give spurious results. Note that \textit{pressure}, as calculated at
present, uses the volume of the simulation cell for the purpose of
conversion and if there is vacuum in any direction the pressure should
not be considered accurate. For this reason \textsc{Conquest}
internally uses values of stress to optimise simulation cells.  

Stress is an extremely useful
quantity: it is used to optimise simulation cell parameters, though
this requires care to converge both the integration grid spacing, and
numbers of k-points. Additionally, it is used in the  NPT ensemble
for molecular dynamics.  

\begin{figure*}[htb]
  \centering
  \includegraphics[width=0.3\linewidth]{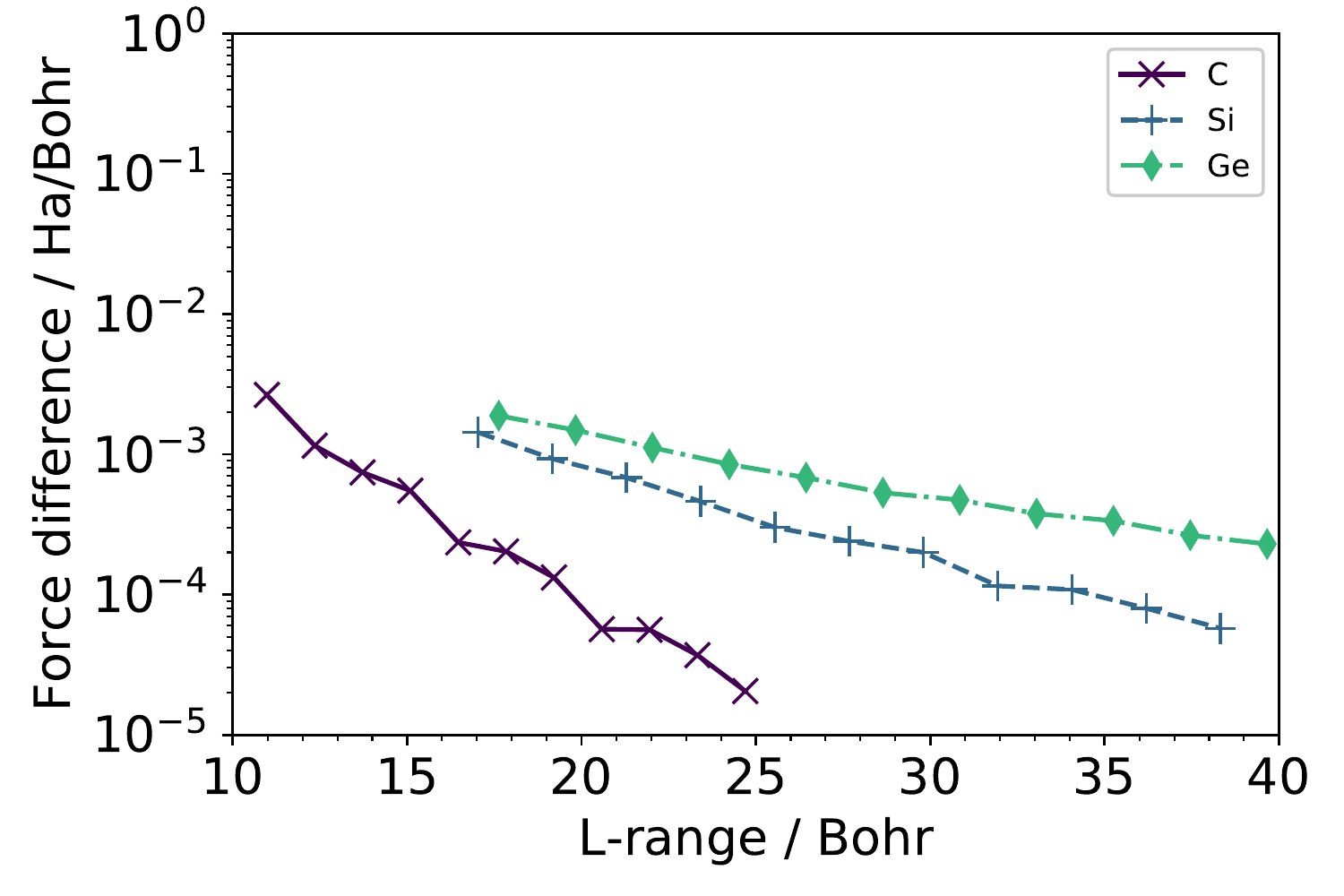}
  \includegraphics[width=0.3\linewidth]{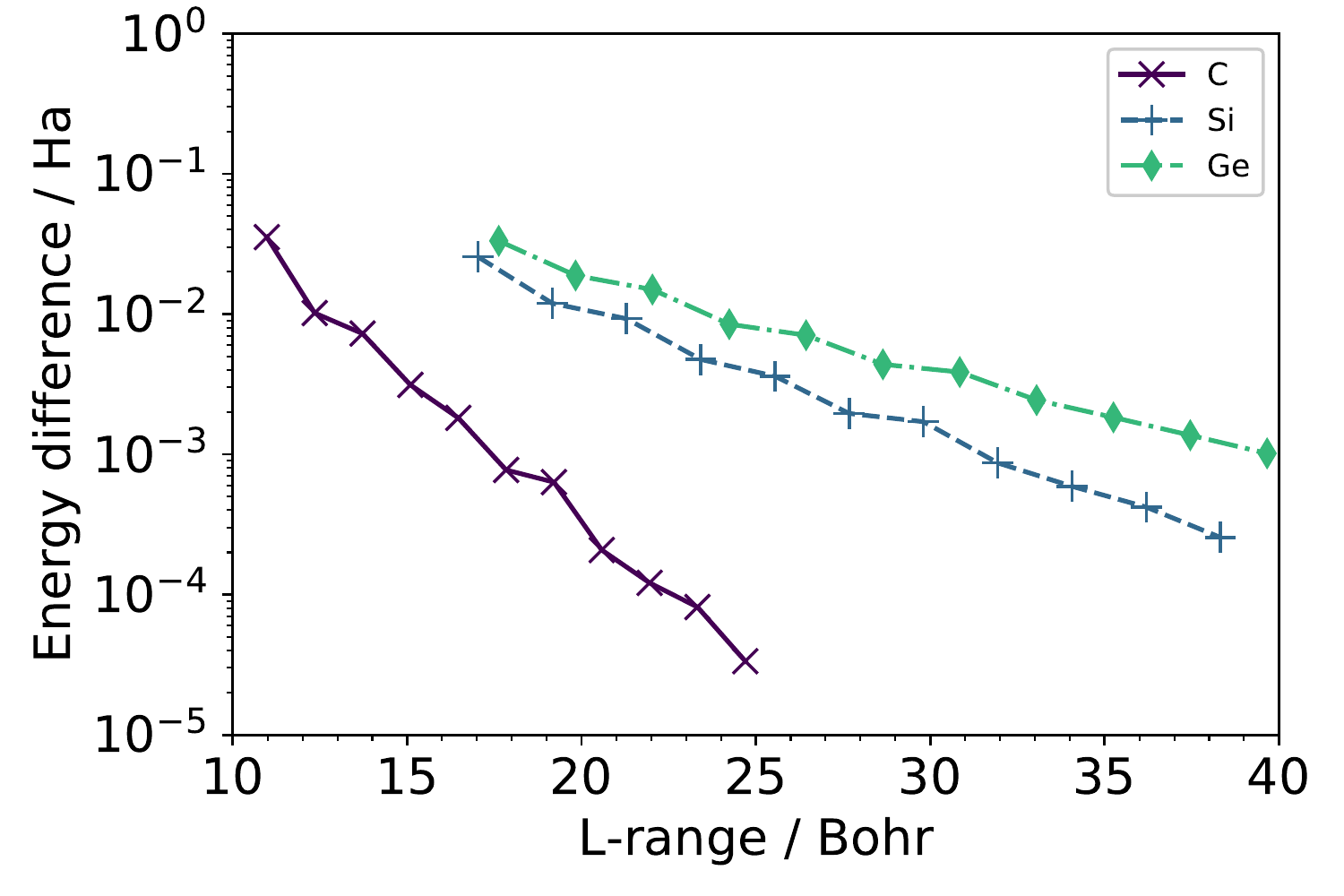}
  \includegraphics[width=0.3\linewidth]{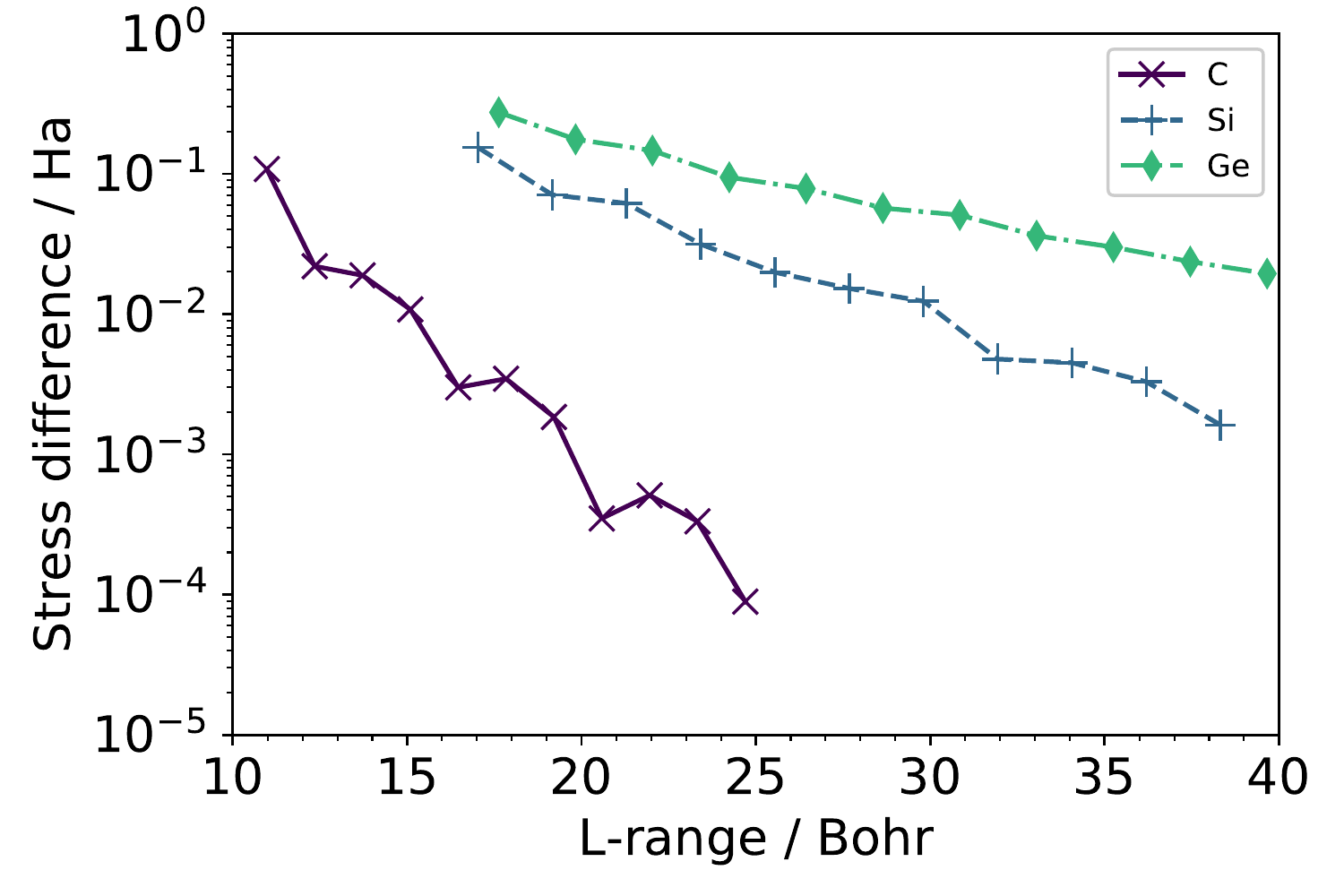}
  \caption{Plots of difference between exact diagonalisation result
    and O(N) result for (a) force; (b) total energy; and (c)
    stress (not normalised by the simulation cell volume) for carbon
    (cross symbols), silicon (plus symbols) and germanium (diamond
    symbols).\label{fig:lrange-conv}}
\end{figure*}

Our implementation of stress is valid for both exact diagonalisation
and linear scaling solvers.  However, we have found that the stress
converges extremely slowly with respect to density matrix truncation.
Figure \ref{fig:lrange-conv} shows the convergence of force
(i.e. energy differences), total energy and stress with density matrix
truncation for three different
elemental semiconductors with very different gaps: carbon, silicon and
germanium.
Calculations were performed on the diamond structure (with a small
perturbation in the case of the force calculation) at the optimal
lattice parameter found using exact diagonalisation, with integration
grid spacing of 0.1 Bohr radii and an 8$\times$8$\times$8
$\Gamma$-centred Monkhorst-Pack grid.  To aid comparison between exact
diagonalisation and linear scaling calculations, we used the simplest
basis set, i.e. single zeta, though this does not change the final
results significantly.  The plots show the difference between the
$\mathcal{O}(N)$ and full diagonalisation results.  The full
diagonalisation results for the stresses were all less than 0.001Ha
(and less than 0.1GPa when converted to a pressure).  For the forces,
the full diagonalisation results were 0.036\,Ha/bohr for C,
0.016\,Ha/bohr for Si and 0.014\,Ha/bohr for Ge.  The total energies
were -47.891\,Ha for C, -33.611\,Ha for Si and -39.589\,Ha for Ge.

The spatial decay of the density matrix is not
analytically described for complex materials, but can be shown to
decay approximately exponentially with gap\cite{Ismail-Beigi1999,He2001,Taraskin2002}:
\begin{equation}
\rho(\mathbf{r},\mathbf{r}^{\prime}) \propto
\exp(-\gamma\mid\mathbf{r}-\mathbf{r}^\prime\mid)
\label{eq:s1}
\end{equation}
We can see in 
Fig.~\ref{fig:lrange-conv} that the rates of convergence of the
different materials with density matrix truncation decrease with
decreasing gap size, as expected. It is notable that the initial
errors are largest for the stresses, and that significant differences
in the stress remain even at very large density matrix ranges.  We
will investigate this fully in a future publication, but we are
confident that this comes from the implicit dependence of energy on
density matrix truncation range, which should be included in a stress
calculation as it will change as the unit cell is changed; however, an
analytic form for this stress is not available.

\subsection{Structure optimisation}

Structural optimisation can be performed using a variety of standard
approaches: the L-BFGS algorithm for atomic optimisation; conjugate
gradients for atomic and simulation cell optimisation; and quenched
molecular dynamics (both in a simple form, and using the FIRE
algorithm\cite{Bitzek:2006yv}).  We note that some form of
preconditioning will become increasingly important as system sizes
increase, and we are planning to implement some recently proposed
preconditioners\cite{Packwood:2016vk,Fernandez-Serra:2003vy}. 

\subsection{\label{sec:MD} Molecular dynamics}
Since the calculated forces are accurate and we can treat large systems,
it is reasonable to expect that we can perform reliable molecular dynamics of
large complex systems using \textsc{Conquest}.
Unfortunately, it is not so easy to realize reliable MD simulations 
with the linear-scaling DFT technique, or with MSSF.
We have two key issues here. 
First, the calculation time for each MD step should be small enough to reach 
a meaningful simulation time.
Second, density matrix should be sufficiently accurate to produce reliable MD simulations. 
During structure optimization, we can refine the accuracy step by
step, without significant penalty. 
In many cases, we need a rather high accuracy only in the late stages of structure optimization. 
On the other hand, for MD simulations, we need to calculate the
density matrix accurately at every step to ensure that the correct
trajectory is followed.
The accuracy of the density matrix depends on the tolerance to which
it is optimised.  Here, the optimized quantities are the auxiliary density matrix $L$ in the 
linear-scaling calculations and PAO coefficients of the support functions
in the MSSF method. Hereafter, we focus the linear-scaling calculations. 
 
\begin{figure}
  \centering
  \includegraphics[width=0.5\linewidth]{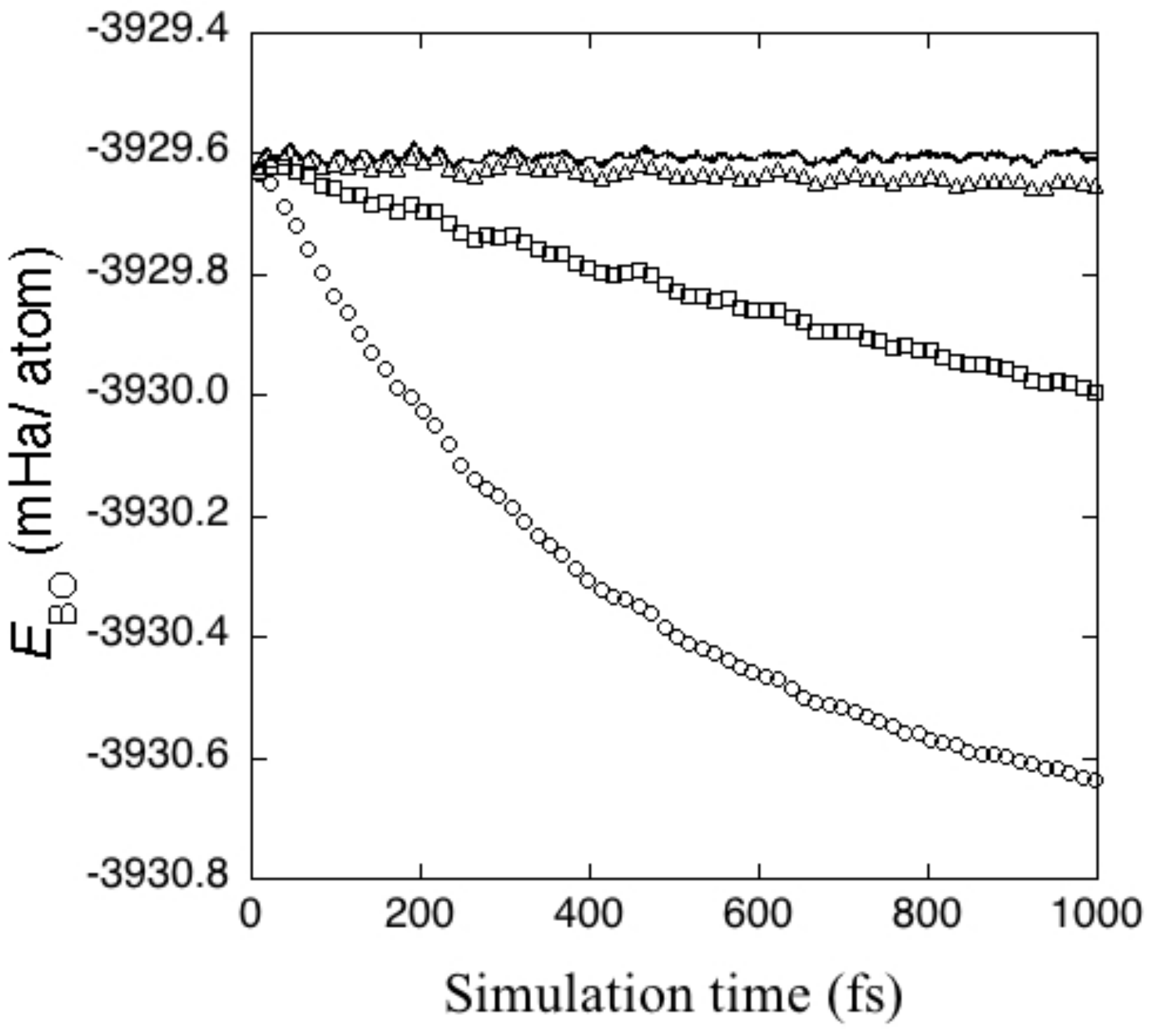}
  \caption{ Time evolution of Born-Oppenheimer total energy
    ($E_{\rm BO}$) obtained by McWeeny initialization at every step
    (solid line) and by reusing the $L$ matrix from the previous step
    for different tolerances (symbols). Symbols indicate tolerances of
    $1.6 \times10^{-5}$ (circles), $1.6 \times10^{-7}$ (squares), and
    $1.6 \times10^{-9}$ (triangles).  Reprinted with permission from
    J. Chem. Theor. Comput. 10, 5419 (2014). Copyright (2014) American Chemical
    Society.}
  \label{fig:MDproblem}
\end{figure}

For efficiency, we need a good initial guess of $L$ matrix, at each MD step,
and the simplest way, which should be efficient, is to use the $L$ matrix optimized at the previous step.
However, as is well known, this breaks the time-reversibility of the dynamics, resulting
in a “drift” in the constant of motion over time\cite{2008.Niklasson.PRL}.
Figure \ref{fig:MDproblem} shows the Born-Oppenheimer total energy ($E_{\rm BO}$), defined 
as the sum of the ionic kinetic energy $T$ and the DFT total energy $V_{\rm BO}$, 
for linear-scaling MD simulations of a 64-atom silicon crystalline
system with different tolerances on the optimisation of the $L$ matrix.
The simulations are performed with the velocity-Verlet integrator with a time step of 0.5 fs 
in a microcanonical ($NVE$) ensemble with initial velocities set so
that the system temperature is 300K.
The symbols in the figure show the time evolution of $E_{\rm BO}$, which should be constant
in reliable NVE-MD simulations. 
The results show that we need a very strict tolerance for stable MD simulations.
Note that, if we use McWeeny initialization at every MD step (shown by solid line 
in Fig. \ref{fig:MDproblem}), $E_{\rm BO}$ is almost constant even if we use a rough tolerance.
But, this leads to a high computational cost at each iteration.

 To solve this problem, \textsc{Conquest} uses the XLBOMD
 method\cite{Niklasson2006, 2008.Niklasson.PRL, Niklasson2017}, with
 the DMM method.  The extended Lagrangian used in \textsc{Conquest}
 is\cite{Arita2014}: 

\begin{eqnarray}
  \mathcal{L}^\mathrm{XBO}\left(\mathbf{X}, \mathbf{\dot{X}}, \mathbf{R},
  \mathbf{\dot{R}}\right) &=& \mathcal{L}^\mathrm{BO}\left(\mathbf{R},
  \mathbf{\dot{R}}\right) + \frac{1}{2}\mu\mathrm{Tr}
  \left[\mathbf{\dot{X}}^2\right] - \nonumber\\
  &&\frac{1}{2}\mu\omega^2\mathrm{Tr}\left[(\mathbf{LS} - \mathbf{X})^2\right],
\end{eqnarray}
where $\mathbf{S}$ is the overlap matrix and $\mathbf{X}$ is a sparse matrix 
associated with $\mathbf{LS}$ rather than $\mathbf{L}$ to maintain the orthogonal metric. 
$\mu$ is the fictitious electronic mass and $\omega$ is the curvature of 
the electronic harmonic potential. If we take the limit $\mu \rightarrow 0$,
$\mathcal{L}^{\rm XBO}$ becomes $\mathcal{L}^{\rm BO}$ and we have equations of motion 
for nuclear positions and $\mathbf{X}$, and for $\mathbf{X}$ 
\begin{equation}
  \mathbf{\ddot{X}} = \omega^2(\mathbf{LS} - \mathbf{X}),
\end{equation}

If we apply the Verlet scheme to calculate $\mathbf{X}$, we have
\begin{equation}
  \mathbf{X}(t+\delta t) = 2\mathbf{X}(t) -\mathbf{X}(t-\delta t) +
  \delta t^2\omega^2\left[\mathbf{L}(t)\mathbf{S}(t)-\mathbf{X}(t)\right] \nonumber,
\end{equation}
i.e. the trajectory of $\mathbf{X}(t)$ is time-reversible, and evolves in a
harmonic potential centred on the ground-state density
$\mathbf{L}(t)\mathbf{S}(t)$. The matrix $\mathbf{XS}^{-1}$ is then used as the
intial guess for the $\mathbf{L}$-matrix.

\begin{figure}
  \centering
  \includegraphics[width=0.6\linewidth]{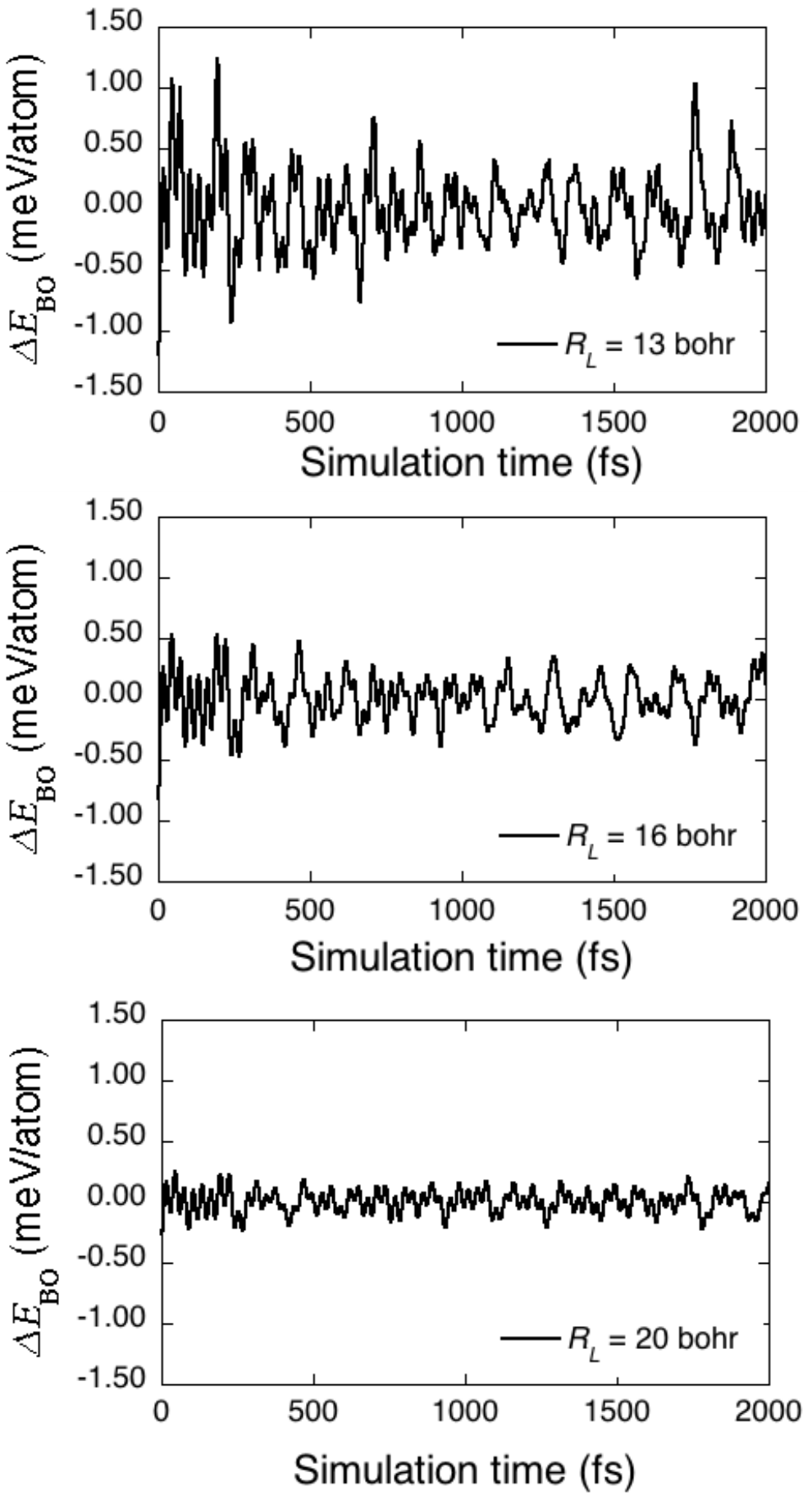}
  \caption{Variation of $E_{\rm BO}$ during the $NVE$ MD simulation
    of crystalline Si, with $R_L$ = 13 bohr (top), 16 bohr (middle),
    and 20 bohr (bottom).  Reprinted with permission from
    J. Chem. Theor. Comput. 10, 5419 (2014). Copyright (2014) American Chemical
    Society.}
  \label{fig:NVE_RL}
\end{figure}

If we use this method, the total energy $E_{\rm BO}$ is stable and the MD trajectories 
do not strongly depend on the tolerance or the range $R_L$ in the
$\mathcal{O}(N)$ calculations\cite{Arita2014}. 
Figure \ref{fig:NVE_RL} shows the variation of the total energy
$E_{\rm BO}$ with simulation time for different values of $R_L$.
The fluctuations in the energy are smaller for larger $R_L$, 
but even with $R_L$ = 13 bohr, the energy drift in $E_{\rm BO}$ is very small, 
meaning that the MD simulation is stable. 

In practice, the $\mathbf{X}$-matrix sometimes moves away from the harmonic
centre over time, increasing the number of SCF iterations required to reach the
ground state over the course of a simulation. To remove this instability, 
the dissipative term, $ a\sum_{m=0}^M c_m\mathbf{X}(t-m\delta t) $,
is included\cite{2009.Niklasson.JCP}. In principle, this dissipation term may break the
time-reversible symmetry, but it is made to have a minimal effect and it is found 
that the MD simulations with the term is stable.

\begin{figure}[h]
  \centering
  \includegraphics[width=0.5\linewidth]{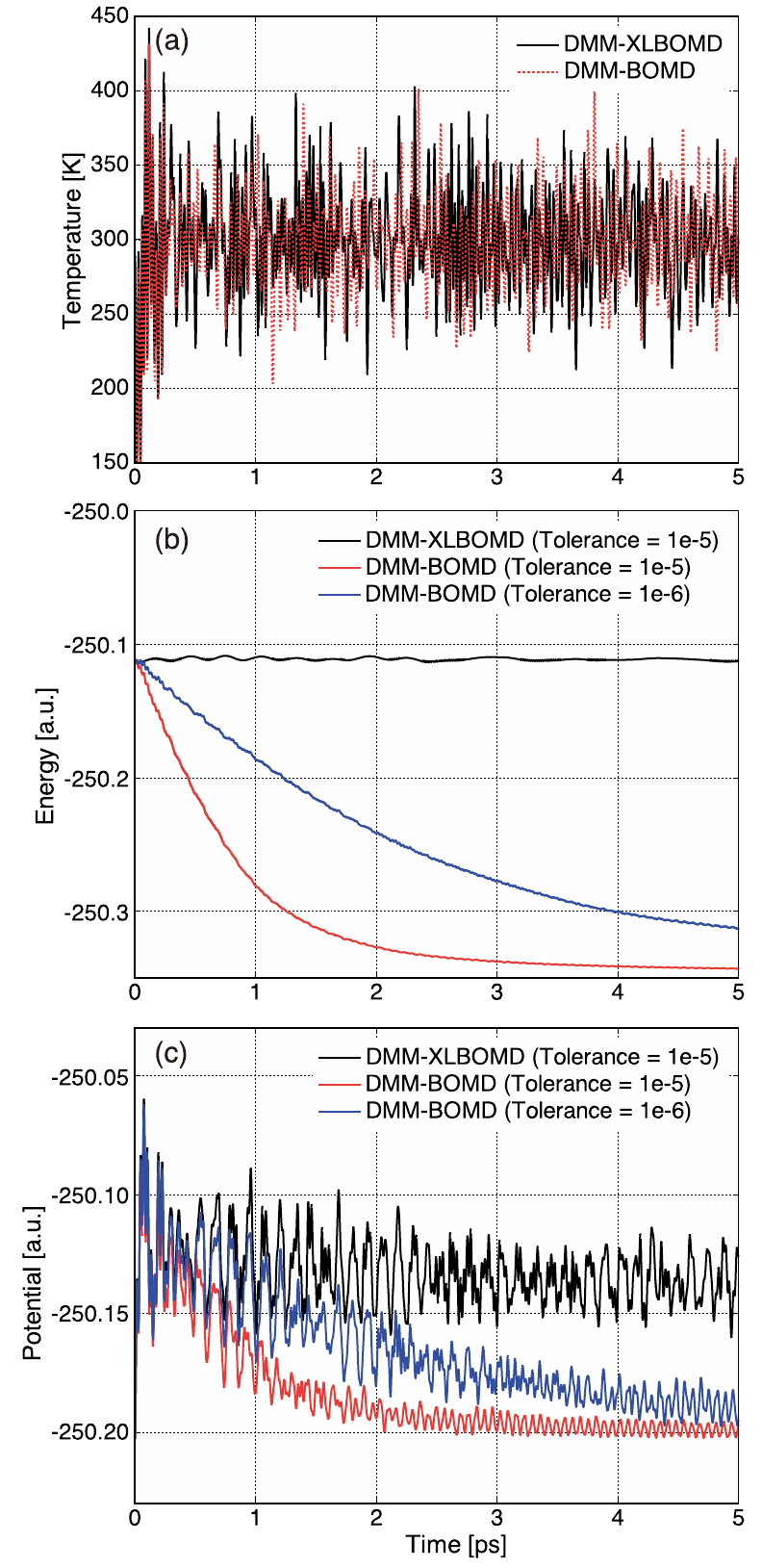}
  \caption{Time evolution of the MD simulations of 64-atom Si
    crystalline systems in the canonical ensemble, for a) Temperature,
    b) constant of motion for the NHC method, and the potential energy
    ($V_{\rm BO}$).  The results with XLBOMD method and those without
    using XLBOMD method, referred to as DMM-BOMD, are compared.
    Reproduced from Ref. \onlinecite{Hirakawa2017} with permission.
    Copyright IOP Publishing.  All rights reserved.}
  \label{fig:NVT}
\end{figure}

Using this XLBOMD + DMM method, we can also treat the canonical ensemble and perform  
constant temperature ($NVT$) MD simulations, for example, using 
the Nos\'e-Hoover chain (NHC) method\cite{Hirakawa2017}. 
The detailed explanation of the integration scheme used for the
canonical ensemble is provided in Sec.~\ref{sec:canon-nvt-ensemble}. 
Figure \ref{fig:NVT} shows the time evolution of the temperature, the constant of motion for the NHC method, 
and the DFT potential energy $V_{\rm BO}$, in NVT-MD simulations of the same crystalline 64-atom 
silicon system at 300K, with and without XLBOMD method.
We find that temperatures are stable and close to 300K in both simulations.
However, we again observe the problem of drift in the constant of motion 
when we do not use the XLBOMD method, while there are no such problems in the XLBOMD+DMM simulations.
More importantly, the profile of $V_{\rm BO}$ is completely different between 
the usual DMM and XLBOMD+DMM MD simulations.  \textsc{Conquest} can
also perform NVT simulations using the SVR (stochastic velocity
rescaling) thermostat\cite{Bussi2009},
which is extremely efficient and provides excellent conservation of
the constant of motion, as described in Sec.~\ref{sec:stoch-veloc-resc}.

Since the stress tensors can be also calculated using
\textsc{Conquest} with the DMM method, as shown in
Sec.~\ref{sec:forces-stresses}, it is also possible to include the
degrees of freedom of the unit cell for $NPT$ simulations with a given
pressure, using the Parrinello-Rahman equations of
motion\cite{Parrinello1981}.  \textsc{Conquest} uses the
Martyna-Tobias-Tuckerman-Klein modification\cite{Martyna1996},
coupling the constant pressure equations of motion to a Nos\'e-Hoover
chain thermostat to recover the NPT ensemble.  The integration scheme
used in the NPT ensemble is also explained in Sec.~\ref{sec:isob-isoth-npt}.

\begin{figure}
  \centering
  \includegraphics[width=0.7\linewidth]{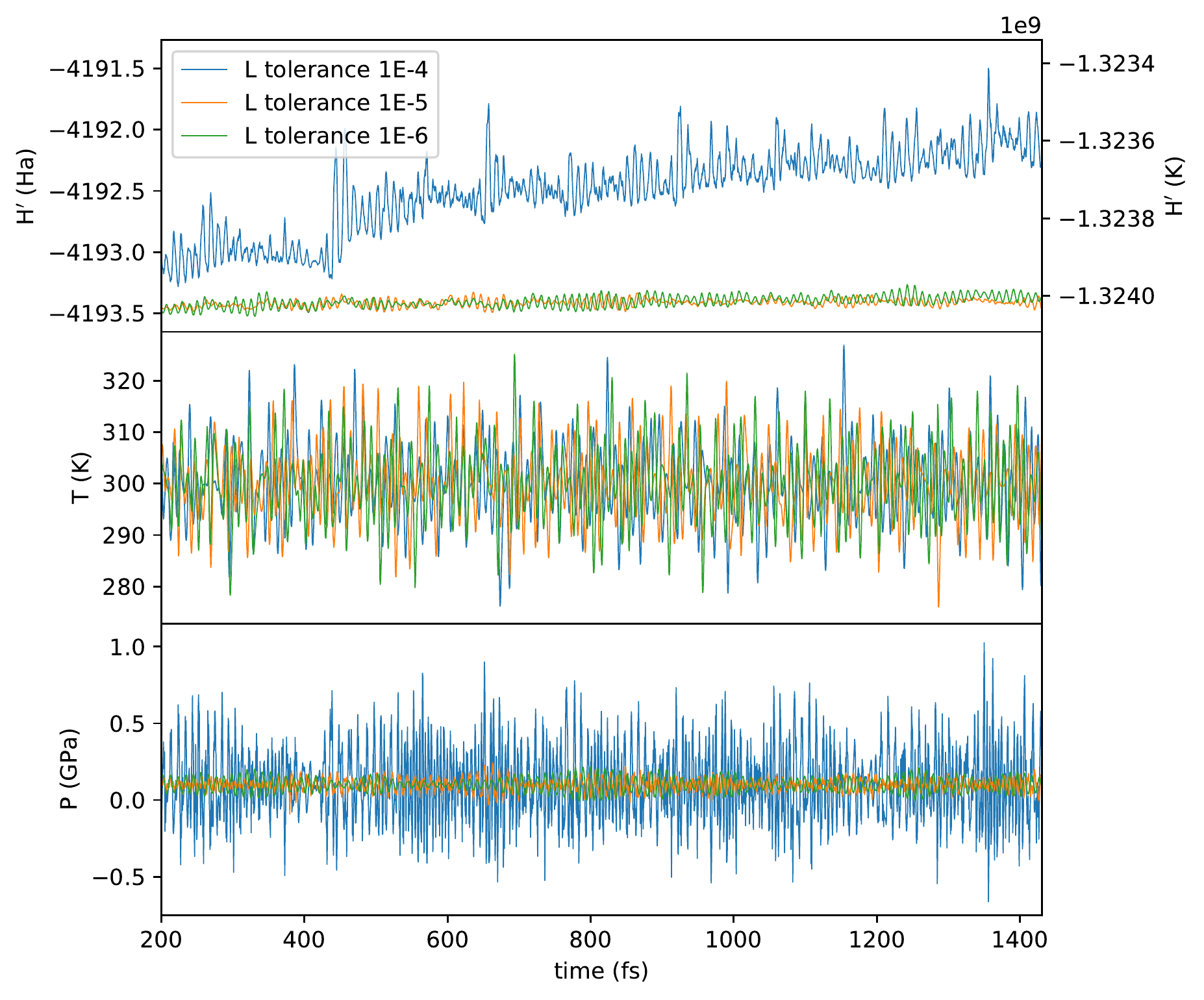}
  \caption{NPT molecular dynamics on 1000 atom bulk silicon system. The three
    lines demonstrate the effect of varying the tolerance applied to
    the optimisation of the energy with respect to the density matrix
    during the $\mathcal{O}(N)$ solution (``L-tolerance'').}
  \label{fig:npt_ordern_ltol}
\end{figure}

This scheme is tested on a bulk crystalline silicon system containing 1,000 atoms, 
and the $\mathcal{O}(N)$ method for finding the electronic ground
state, as shown in Fig.~\ref{fig:npt_ordern_ltol}. A minimal basis set (SZ) was employed, together with a
grid cutoff of 100 Ha and the PBE exchange-correlation functional. The
extended-Lagrangian scheme described above was used, with a velocity Verlet
integrator for the $X$ matrix and 5th order dissipation. The system was
equilibrated using a Berendsen-type weak coupling thermostat and barostat, at a
temperature of 300 K and pressure of 0.1 GPa. The cell volume was allowed to
vary, but constrained to be cubic. An integration time step of 0.5 fs was used,
with a 5th-order Yoshida-Suzuki integration scheme, and thermostat and barostat
coupling time periods of 15 fs and 160 fs respectively. An \textit{ad hoc} drag
was applied to the barostat, reducing the velocities of the cell and its
Nos\'e-Hoover thermostats by 5\% each time-step. This was found to improve the
stability, preventing the amplification of ``ringing'' of the barostat, with a
minimal impact on energy conservation.

It can be seen that in order to achieve good energy conservation, the
L-tolerance lower than $10^{-5}$, with a significant drift in the conserved
quantity occuring at looser tolerances; without the XL-BOMD scheme,
the tolerance required would be much tighter. We note that the NPT integrator is considerably more
sensitive to the time step due to coupling between the thermostat and barostat
degrees of freedom, and that in this case a time step of 1.0 fs also resulted in
a significant energy drift, though we are seeking to alleviate this sensitivity.

\section{\label{sec:perform} Performance}

Here we demonstrate the performance of \textsc{Conquest} showing two
examples, one for the MSSFs with diagonalisation and another for the
$\mathcal{O}(N)$ calculations. 

\subsection{\label{subsec:performMSSF} Performance of MSSF}
\begin{figure*}
  \centering
  \includegraphics{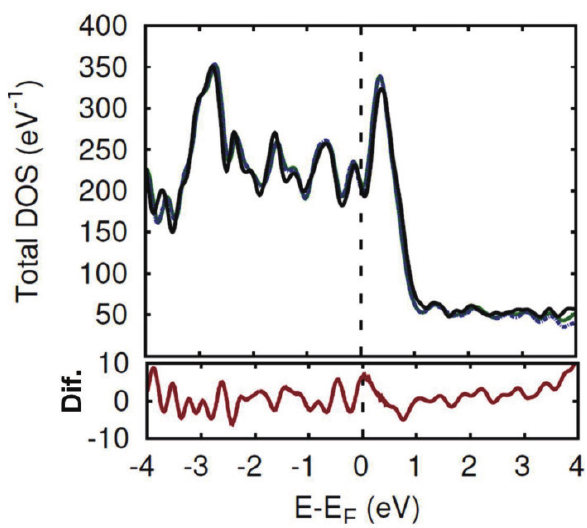}
  \caption{Density of states of m14 graphene/Rh(111) calculated with
    plane-waves (black), PAOs (green) and MSSFs (blue). The red line in
    the lower panel represents the difference between the DOS
    calculated with PAOs and
    MSSFs.  Reproduced from Ref.~\protect\onlinecite{RomeroMuiz2018})
    with permission. Copyright IOP Publishing.  All rights
    reserved.\label{fig:PerformanceMSSF1}}
\end{figure*}

A recent study on the graphene/Rh(111) interface \cite{RomeroMuiz2018}
showcases both the accuracy and efficiency of the MSSFs.  This study
used large basis sets of PAOs contracted to a minimal size using the
MSSF formalism, i.e., 15 and 22 PAOs of rhodium and carbon atoms are
contracted to 6 and 4 MSSFs, respectively.  In
Ref.~\onlinecite{RomeroMuiz2018}, it was demonstrated that the PAOs and
MSSFs show comparable accuracy with plane-waves for the electronic and atomic structure
of graphene/Rh(111), as shown in Fig.~\ref{fig:PerformanceMSSF1}.  The
accuracy of PAOs has been further investigated in Ref.~\onlinecite{Bowler:2019fv}.

\begin{figure*}
  \centering
  \includegraphics{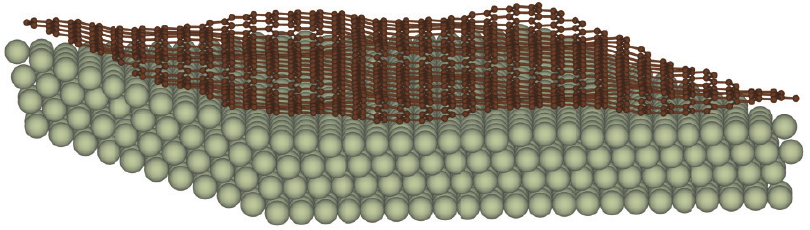}
  \caption{Atomic structure of graphene/Rh(111) system (3,088 atoms).\label{fig:PerformanceMSSF2}}
\end{figure*}

This study then demonstrates the great reduction in computational
effort by using MSSFs.  Table~\ref{table:PerformanceMSSF1} shows the
computational times of a SCF step for the graphene/Rh(111) systems
consisting of 1544 and 3088 atoms (shown in Fig.\ref{fig:PerformanceMSSF2}).
MSSF clearly require more computational time for matrix
construction than the PAOs, which comes from the calculations of the linear
combination coefficients, as explained in
Sec.~\ref{sec:methods-elec-struc}. On the other hand, the time to
diagonalise the electronic Hamiltonian is reduced significantly by
using MSSFs, because diagonalisation time scales cubically with the
number of support functions. For the 1,544-atom system, the total time,
i.e., the summation time of matrix construction and diagonalisation,
is reduced by a factor of $\approx 3$, from 1,256.9 seconds to 439.6
seconds. For the 3,088-atom systems, when using 108 processes, the
total time is reduced by a factor of $\approx 18$, from 37,803.5
seconds to 2,156.3 seconds, which indicates that the use of MSSFs
becomes more efficient as systems become larger. Comparing the time
for the matrix construction for the 1,544 atoms with 432 MPI processes
and that for the 3,088 atoms with 864 MPI processes, i.e., when both
the system size and the number of processes are doubled, the times are
very close to each other, which indicates the construction of the
MSSFs is $\mathcal{O}(N)$ and parallelized ideally.

\begin{table}
  \begin{tabular}{lrrrrr}
  \hline
                      & \multicolumn{2}{c}{1544 atoms} & \multicolumn{3}{c}{3088 atoms} \\ 
                      &   PAO   &  MSSF  &   PAO   & MSSF   &  MSSF  \\
    \hline
  No. of MPI process  &   432   &   432  &   108   &  108   &  864   \\
                                                                       \\
  Time [sec.]         &         &        &         &        &        \\
  matrix construction &    64.3 &  400.4 &   155.7 & 1455.4 &  405.9 \\
  diagonalisation     &  1192.5 &   39.2 & 37647.7 &  700.8 &  165.9 \\
  total\footnote{Summation of matrix construction and
    diagonalisation.}
                      &  1256.9 &  439.6 & 37803.5 & 2156.3 &  571.8 \\
  \hline  \end{tabular}
  \caption{Computational time for self-consistent-field calculation
    step with PAOs and MSSFs for graphene/Rh(111) performed on the
    supercomputer SGI ICE X in
    NIMS; data from Ref.~\protect\onlinecite{RomeroMuiz2018}.\label{table:PerformanceMSSF1}}
\end{table}

\subsection{\label{subsec:performOrderN} Performance of $\mathcal{O}(N)$ calculations on massively parallel computers}
\begin{figure}
  \centering
  \includegraphics[width=\linewidth]{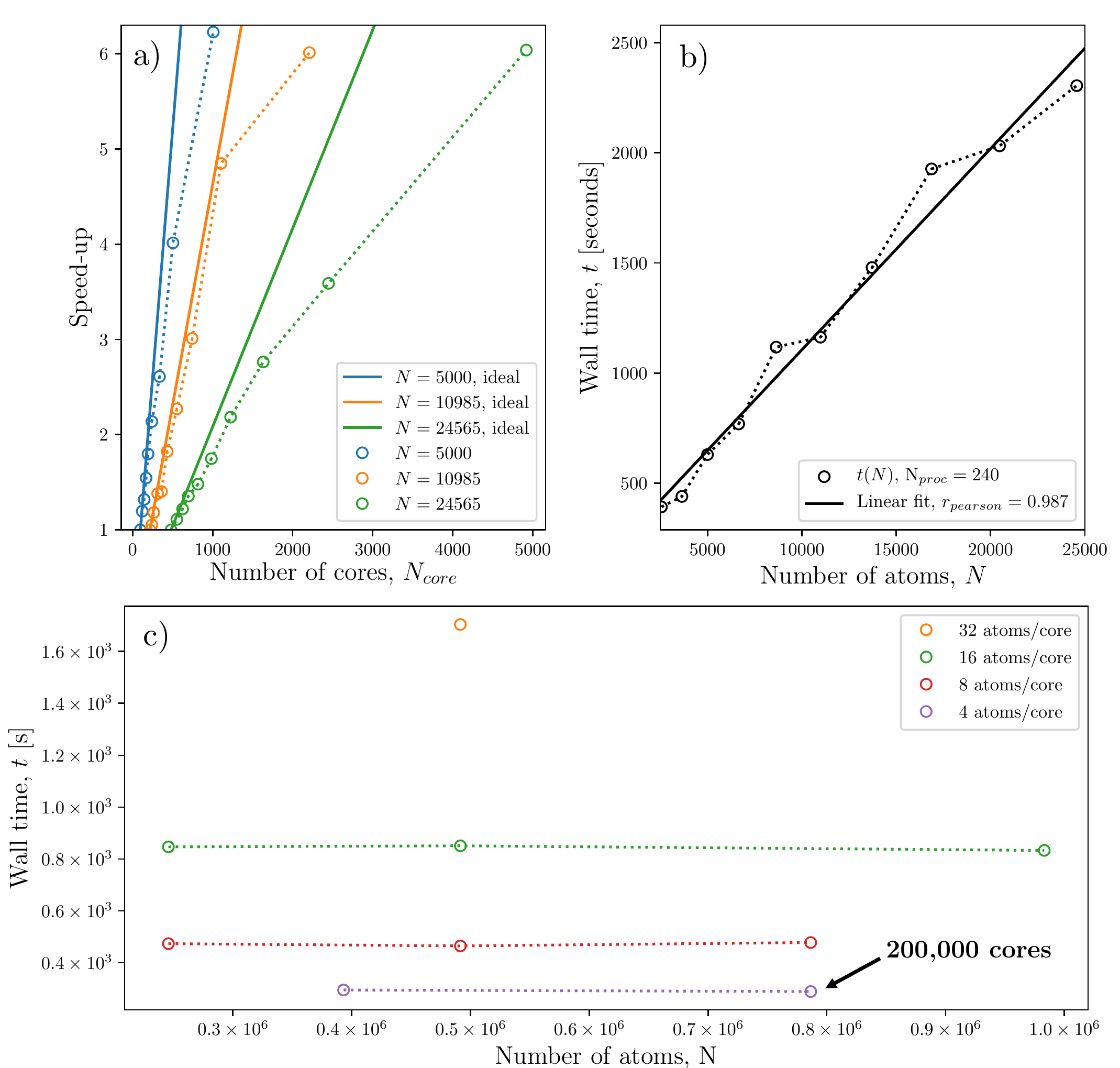}
  \caption{Scaling performance for \textsc{Conquest} on the K-computer
    and ARCHER. (a) Strong scaling on the UK national supercomputer,
    ARCHER, up to 4,920 processors (from 50 atoms/core up to 5
    atoms/core).  Calculations are performed on bulk
    PbTiO$_3$ with an $L_{\text{range}}$ of 14$a_0$ and a SZ basis. (b)
    Demonstration of the scaling of the $\mathcal{O}(N)$ algorithm on
    ARCHER for the same material system as (a). (c) Weak scaling on the
    K-computer up to 1,000,000 atoms for bulk Si.\label{fig:Scaling}}
\end{figure}

The performance of \textsc{Conquest} on the Japanese, Fujitsu-made K-computer
is of real significance \cite{Arita2014a}.  This computer once topped
the TOP500 list \cite{Meuer2012} (June and November 2011) and 8 years
later \textit{still featured} on the list in 20th place (November
2019) due to its impressive peak performance of 11,280.4 TFlops/s from
its 705,024 physical cores.  \textsc{Conquest} was found to display almost
ideal parallel efficiency, as shown in Fig.~\ref{fig:Scaling}c, utilising up to 200,000 
physical cores\cite{Arita2014a} on systems up to 2 million atoms
\cite{Bowler2010}.  (At present, there is no dynamic
  fault-tolerance built in to \textsc{Conquest},  
to account for failure of nodes during a run; however, the frequency 
with which restart files are written can be controlled at a fine-grained
level, which makes recovery from a crash easy.) Using crystalline silicon systems as a
benchmark, it was demonstrated that in the $\mathcal{O}(N)$ mode of
operation that both strong scaling (the wall time for a fixed number
of atoms, increasing the physical core count) and weak scaling (the
wall time for a fixed number of atoms/physical core, increasing the
number of atoms) performs very well.  Specifically, for strong scaling
it is found that performance is good should the number of atoms/core
be $\geq 4$ but for weak scaling, the performance is close to
\textit{perfect} for any given number of atoms per core all they way
up to 2,000,000 atoms \cite{Bowler2010}.  Strong scaling has also been
tested on the UK national supercomputer ARCHER, a Cray XC30 MPP system
(Figure \ref{fig:Scaling}a).  This demonstrates also the high
efficiency of the code until about  5 atoms/core.
Going to fewer atoms/core than this starts
to significantly impact the performance of \textsc{Conquest}; for this
particular test, more than 50 atoms/core was feasible, but for
more stringent tests, it would require large amounts of memory.
When testing the
scalability of the $\mathcal{O}(N)$ algorithm itself (Figure
\ref{fig:Scaling}b), we see that we achieve near-perfect linear
scaling with system size even in the range of 2560-24565 atoms.

\section{\label{sec:App} Applications}

There are a multitude of physical systems to which \textsc{Conquest} has
been applied.  Studies using both exact diagonalisation (with and
without the use of MSSFs) and the $\mathcal{O}(N)$ mode of operation
have all been exploited in large-scale structural relaxations and
molecular dynamics.  In the solid state, the code has been used to
study the properties of nanowires \cite{ORourke2018,Kumarasinghe2019},
Ge hut clusters on Si (001) surfaces \cite{Miyazaki2008}, charge
transport properties\cite{Li2019}, interfaces between graphene with
metals \cite{RomeroMuiz2018} and ferroelectric domain morphologies in
perovskite oxide heterostructures.  The code has also been applied to
complex biological systems including hydrated DNA \cite{Otsuka2008,
  Otsuka2016} and gramicidin-A \cite{Todorovi2013}.  It is the purpose
of this section to outline some of these studies and to
suggest areas that the code could find new applications.

\subsection{\label{subsec:AppSiGe} Nanoscale Ge/Si systems}
One of the most important targets for large-scale DFT study is
nano-structured semiconductors.  Among them, Ge/Si systems have many
attractive properties as a candidate for next-generation devices.
Heteroepitaxy and strained growth in Ge/Si systems can be used as
important techniques to control the structures and to explore new
favorable properties.

\textsc{Conquest} was first applied to study the stability of Ge
three-dimensional islands on Si substrate, called hut clusters, made
of four equivalent Ge(105) facets.  Experimentally, this 3D
structure appears when the coverage of Ge atoms becomes large, after
the formation of a two-dimensional (2D) structure with
defects\cite{Mo1990}.  Here, \textsc{Conquest}
calculations were performed with LDA and non-selfconsistent mode using
a minimal basis set (SZ), whose accuracy were througly investigated
for Ge/Si systems\cite{Miyazaki2007}.  The stability of the 3D structure in the
heteroepitaxy systems is usually determined by the competition between
the energy gain of the strain relief by the 3D structure and the
energy loss due to the increase of the surface area by the formation
of facets in the 3D structures.  However, the Ge/Si(001) system has the
unique property that the strained Ge(105) facet is more stable than
the strained Ge(001) surface\cite{Hashimoto2005, Raiteri2002} even
accounting for the increase in surface area.  Thus, in order to clarify the
stability of the 3D structure, it is necessary to include the effects
of the edges between the facets and the finite area of the actual
facets. For this, we need to treat the actual size of the hut clusters
with a Si substrate.  Standard DFT methods cannot treat the 3D
structure with a size similar to experiments, but it is possible
using \textsc{Conquest} with structure optimization.  In the early
study with \textsc{Conquest}, the total energies of systems having the same
coverage of Ge atoms were compared between the 2D and 3D structures
and it was found that the 3D structure becomes more stable when the
coverage of Ge is larger than 2.7 monolayers\cite{Miyazaki2008}.  This
is close to the minimum coverage showing the transition from 2D to 3D
growth in experiments, supporting the high accuracy of the present DFT
method.

\begin{figure}
  \centering
  \includegraphics[width=\linewidth]{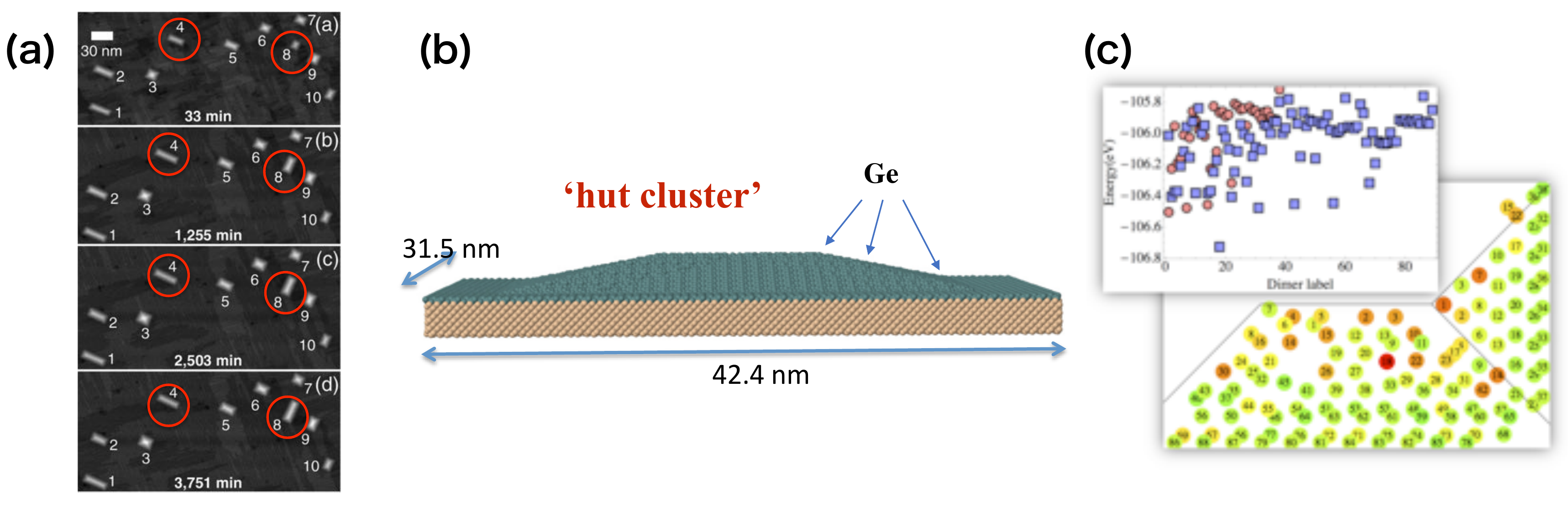}
  \caption{Linear-scaling DFT study of Ge 3D structure on Si substrate
    using \textsc{Conquest}.  (a) Experimental observation showing that
    Ge hut clusters grow, under certain conditions, by
    increasing the length of the longer side while keeping the width
    (shorter side) unchanged. (Reprinted from a figure\protect\cite{McKay2008} in
    Phys. Rev. Lett. 101, 216104 (2008) with permission.  Copyright (2008) by
    the American Physical Society.) (b) The optimized structure of the
    largest structural model for Ge hut cluster on Si substrate, which
    contains about 200,000 atoms.  (c) Adsorption energy map of single
    Ge dimers adsorbed on the \{105\} facets. Disks show the position of
    a given dimer on a facet projected on the x-y plane.  Dimers are
    labelled according to their height. Adsorption energy of dimers
    increases from red to green.  top: Energy values of single Ge
    dimers on small (circles) and large (rectangles)
    facets. (Parts b and c reproduced from 
    Ref.~\protect\onlinecite{Miyazaki2018} with permission.  Copyright IOP Publishing.  All rights
    reserved.)\label{fig:GeHut}}
\end{figure}

Further studies considered the stability of a single Ge dimer adsorbed at various
sites on the facets\cite{Arapan2015}.  This study
aimed to clarify the initial process during the formation of a new
facet layer.  Experimentally, it has been reported that elongated hut
clusters tend to grow, under certain growth conditions, by
increasing the length of the longer side while keeping the width
(shorter side) unchanged (See Fig. \ref{fig:GeHut}(a)
\cite{McKay2008}).  The detailed mechanism underlying the growth
of new facet layers is extremely difficult to obtain from experiments,
since the complete facet is formed rapidly.  We expect large-scale DFT
calculations to play a significant role in clarifying these processes.
By performing structure optimization for more than 100 different sites
for the adsorption site of a single Ge dimer, as shown in Figure
\ref{fig:GeHut} (c), it was suggested that the top or the edges of the
facets are the most preferable sites, and higher positions are more
stable than lower ones.  This kind of study is now possible with
\textsc{Conquest} using a parallel supercomputer.  The largest system
in this study contains about 200,000 atoms, whose structure is shown
in Fig. \ref{fig:GeHut}(b).  Together with the study of double and
triple dimer adsorptions, it was concluded that the new layer of facet
is very likely to grow from top to bottom.

\begin{figure}
  \centering
  \includegraphics[width=\linewidth]{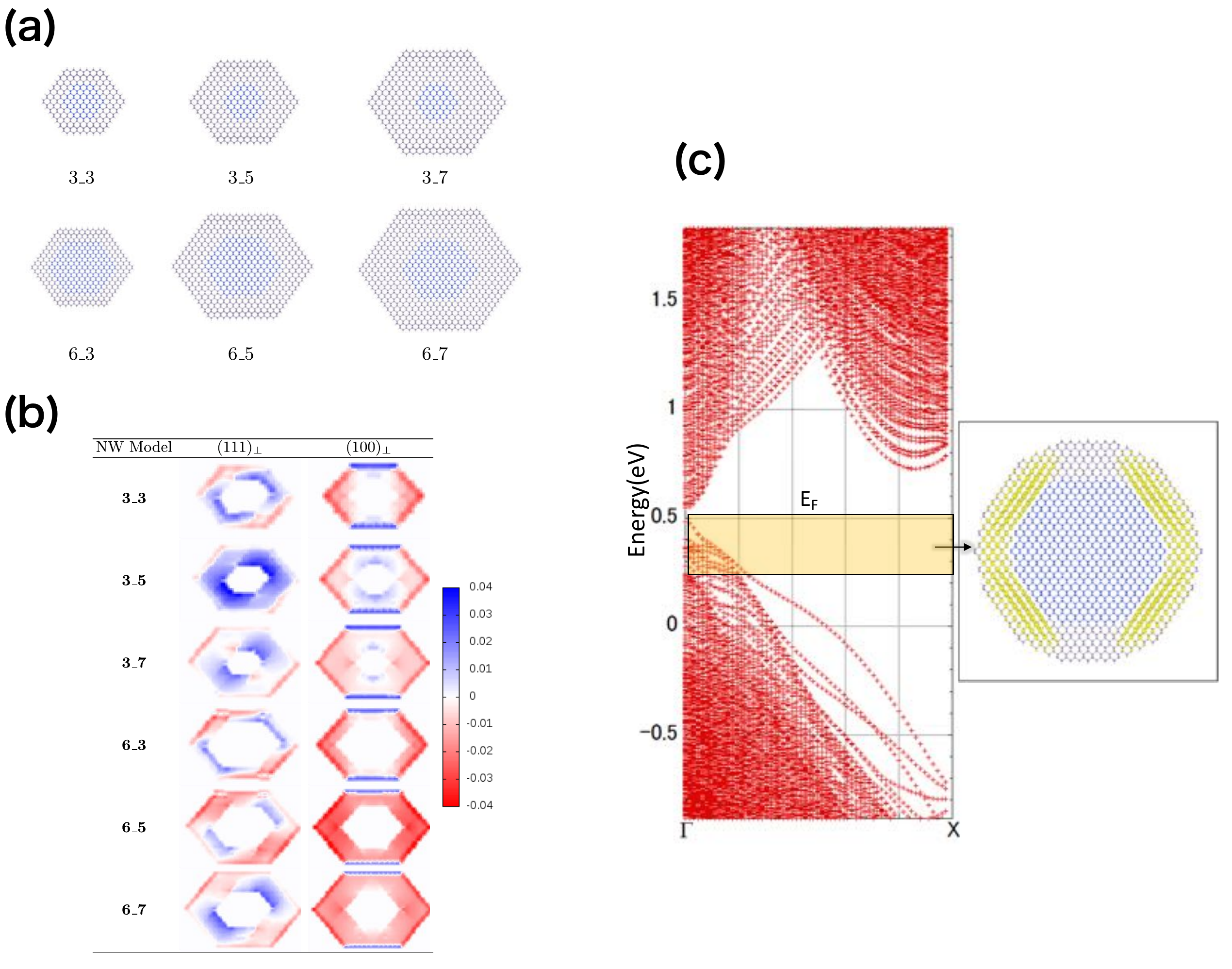}
  \caption{ (a) Structural models used in the study of Si/Ge nanowires
    along \textless 110 \textgreater direction, labelled C\_S where
    the index C represents the number of layers in the core and S the
    surface.  Shell thickness increases left to right, and core
    thickness top to bottom.  (b) Average bond strain map for the
    cross-section of the Ge shell of the SiGe-NWs.  Maps for the bonds
    along different directions are shown, with extension illustrated
    in blue and compression in red.  (c) Band structure of the
    circular Si/Ge core-shell nanowire and the charge density
    constructed from the occupied orbitals near the Fermi level (in
    the range shown in the band structure). (a) and (b) are reproduced
    from Ref.~\protect\onlinecite{ORourke2018} with permission. (c) is
    reproduced from Ref.~\protect\onlinecite{Miyazaki2018} with
    permission.  All parts copyright IOP Publishing.  All rights
    reserved.\label{fig:NW}}
\end{figure}

Recently, \textsc{Conquest} was also applied to study Si/Ge and Ge/Si
core-shell nanowires, using SZP basis set with self-consistency.
Semiconductor nanowires are promising candidates for the
next-generation vertical-type transistors\cite{Thelander2006}, and
have been extensively studied both experimentally and theoretically.
The core-shell type nanowires have many interesting and attractive
properties\cite{Lu2005,Xiang2006,Fukata2012} for next-generation
electronics.  All of these properties, however, will depend strongly
on the size of core and shell. 
Using \textsc{Conquest} with the $\mathcal{O}(N)$ method, strain distributions
were calculated for nanowires with different sizes, shown in
Fig. \ref{fig:NW}(a) up to experimentally accessible sizes.
These are hexagonal Si/Ge core-shell nanowires
along $\langle 110 \rangle$ direction, and with numbers of atoms
ranging from 612 to 2,404.  The strain distributions in the core region
of these nanowires are shown in Fig. \ref{fig:NW}(b).  We can see that
the strain is distributed anisotropically, depending on the direction
of the bonds, and that large
variations of strains exist in the interface and surface regions.

The structure of a more circular Si/Ge core-shell nanowire was also
investigated and its band structure were calculated with the
Sakurai-Sugiura (SS) method explained in
Sec.~\ref{sec:electr-struct-large}, using the optimized structure and
the self-consistent charge density obtained by $\mathcal{O}(N)$ calculations.  The
occupied eigenstates near the Fermi level were also calculated and
are shown in Fig. \ref{fig:NW}(c).  We can clearly see that the
distribution is anisotropic and localized in the Ge-shell region.  The
effect of arsenic doping and its dependence on the doping sites in the
Si nanowires were also recently reported\cite{Kumarasinghe2019} using
a rather high quality basis set (TZTP), with the MSSF method.

\subsection{\label{subsec:AppPerovskite} PbTiO$_3$ films on SrTiO$_3$ substrates}
\begin{figure}
  \centering
  \includegraphics[width=\linewidth]{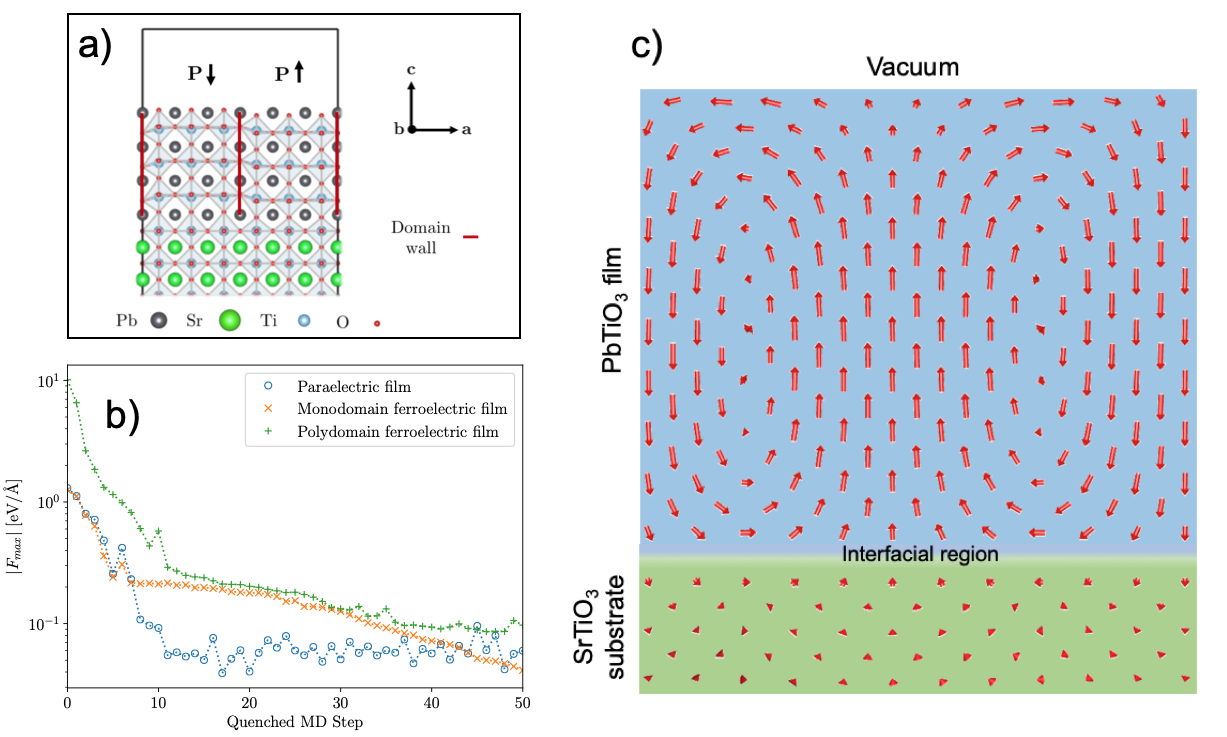}
  \caption{The results of structural relaxation calculations with
    \textsc{Conquest}. (a) The initial geometry used to study
    ferroelectric flux closure domains in PbTiO$_3$ films on SrTiO$_3$
    substrates. This example shows a three unit cell deep film with
    domain period of six unit cells. (b) The evolution the magnitude of
    maximum force on any atom  for the first 50 quenched molecular
    dynamics steps for three different film configurations. (c) The
    local polarisation vector field of a 9 unit cell deep PbTiO$_3$ film
    (2,088 atoms) on a SrTiO$_3$ substrate.\label{fig:structRelax}}
\end{figure}

Studies of the perovskite oxides can also make good use of large-scale
electronic structure calculations.  \textsc{Conquest}
can be used to study large supercells of technologically relevant
piezoelectric alloys like PbZr$_x$Ti$_{1-x}$O$_3$, where approximations
designed to circumvent the need for large supercell calculations (like
the virtual crystal approximation) are unable to quantify local
structural distortions \cite{Baker2019}.  The study of ferroelectric
domains in thin films is another problem requiring large-scale
electronic structure calculations and \textit{accurate} structural
relaxations.  Using the MSSF method and a large basis set of PAOs
(DZDP), the nature of the ferroelectric flux closure domains in thin
PbTiO$_3$ films on SrTiO$_3$ substrates was revealed.  Using the
initial geometry displayed in figure \ref{fig:structRelax}a, we were
able to relax the system, to a stringent 0.01 eV/\AA\ force
tolerance using quenched molecular dynamics.  The force reduction for
the first 50 steps is shown in figure \ref{fig:structRelax}b.  Figure
\ref{fig:structRelax}c shows the local polarisation vector field of a
nine unit cell deep film.  Such a field is calculated using the
relaxed structure, the deviation in displacement from high symmetry
sites and the Born effective charge tensors \cite{Meyer2002}.

\subsection{\label{subsec:AppBio} Biological sysetms}
\begin{figure}
  \centering
  \includegraphics[width=\linewidth]{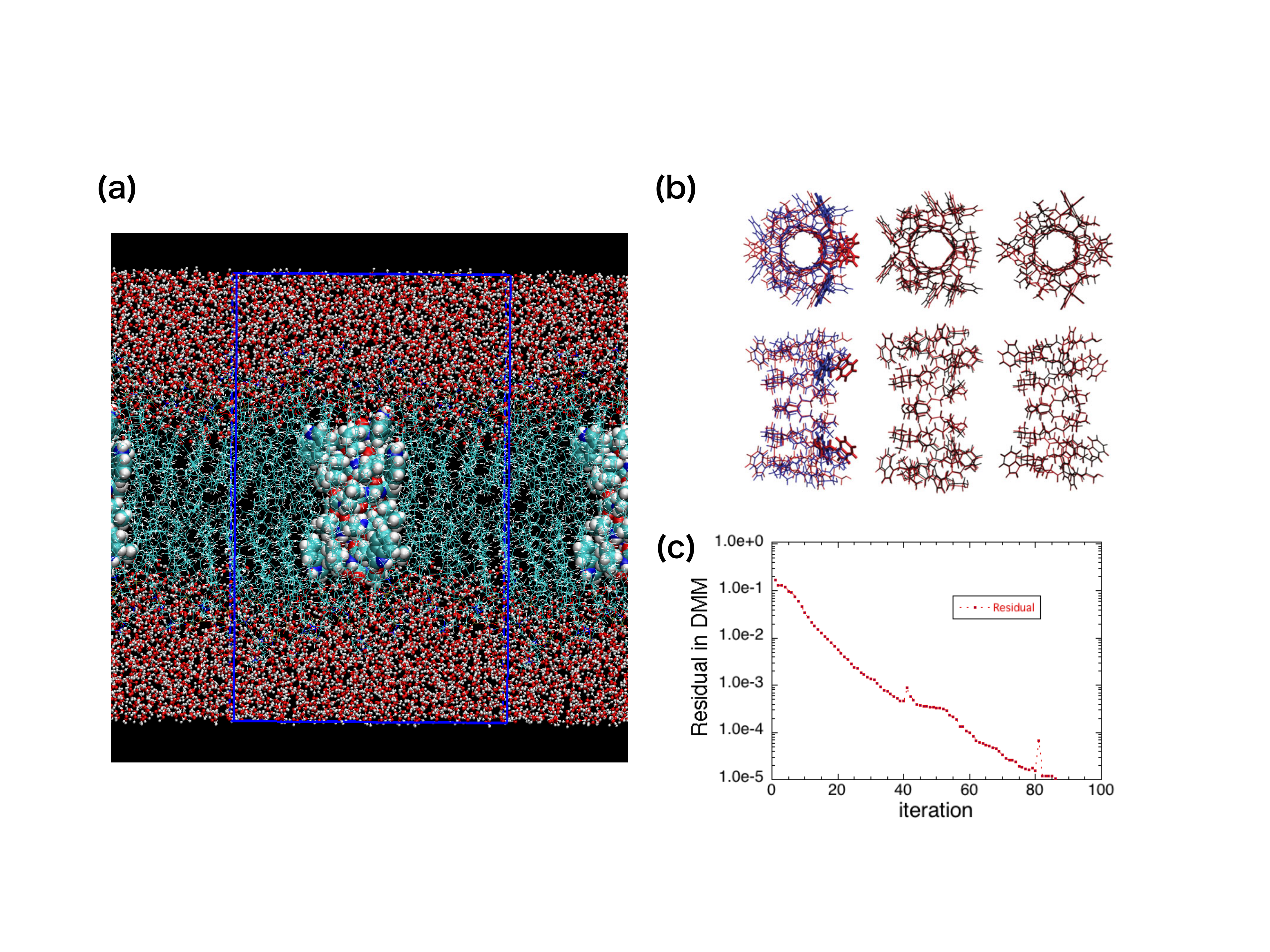}
  \caption{(a) Structure model for the ion channel gramicidin A
    embedded in DMPC lipid bilayers, sandwitched with bulk water
    regions.  (b) Left: Comparison of 1MAG (in blue) and 1JNO (in red)
    experimental structures.  Middle and right: Optimized (red/light)
    and initial (black/dark) structures of isolated gA molecule
    starting from 1MAG (middle) or 1JNO (right) models.  (c) Change of
    the residual during the density matrix minimization (DMM) step in
    the $\mathcal{O}(N)$ calculations of the gA system shown in (a).
    In this calculation, the search direction in the minimization was
    reset at every 40 iterations.
    (a) and (b) are reproduced from
    Ref.~\protect\onlinecite{Todorovi2013} with permission.\label{fig:gA}}
\end{figure}

Complex biological systems are one of the most important targets for
large scale DFT simulations\cite{Cole:2016im}.  \textsc{Conquest} has
been already applied to several biological systems, such as hydrated
DNA\cite{Otsuka2008, Otsuka2016}, dihydrofolate reductase
(DHFR)\cite{Gillan2007}, and the gramicidin A (gA) ion
channel\cite{Todorovi2013} systems.  In the study of gA system, the
optimized structure of the isolated gA molecule, shown in
Fig.~\ref{fig:gA}(b), was first calculated for the two previously
reported structural models, 1MAG and 1JNO.  The electronic structure
of the gA molecule was also analyzed and it was concluded that the
side chains of gA does not affect the electrostatic potential in the
pore of gA.  This kind of study of the isolated gA molecule cannot explain the
selectivity of the ion permeation in the gA system, and it suggests
the importance of simulating the system in the channel environment. We
should treat the gA molecule in lipid bilayers sandwiched by bulk
water regions, as is shown in Fig.~\ref{fig:gA}(a).  Using
\textsc{Conquest}, we can perform stable self-consistent DFT
calculations of such a complex system made of 17,102 atoms, having a
rather irregular charge distribution.  Figure~\ref{fig:gA}(c) shows
that the the density matrix minimization, for a SZP basis set, of this system is
robust.  It is important to note that, for stability
in the self-consistency process, we need to update the charge density
as well as the density matrix at each step in the calculation.  More
detailed information of this large-scale DFT study on the gA system
will be reported in the future.

\subsection{\label{subsec:MD} Large-scale MD simulations with \textsc{Conquest}}
In Sec. \ref{subsec:AppSiGe}, we introduced energetically stable
structures of the perfect epitaxial models for Si/Ge core-shell
nanowires.  However, it is also important to investigate defects
or Si-Ge intermixing at the interface for the actual nanowires.  In
addition, we sometimes need to clarify the thermodynamic stability or
the dynamical processes.  In such cases, molecular dynamics
simulations based on DFT (DFT-MD) are useful and important.  Using the
XLBOMD + DMM method, explained in Sec. \ref{sec:meth-moving-atoms}(c),
we are now able to do practical and reliable self-consistent DFT-MD
simulations of very large systems.  The MD simulations in this
section used SZ (NWs, SiO$_{2}$) and SZP (hydrated DNA) basis sets.

\begin{figure}[h]
  \centering
  \includegraphics[width=\linewidth]{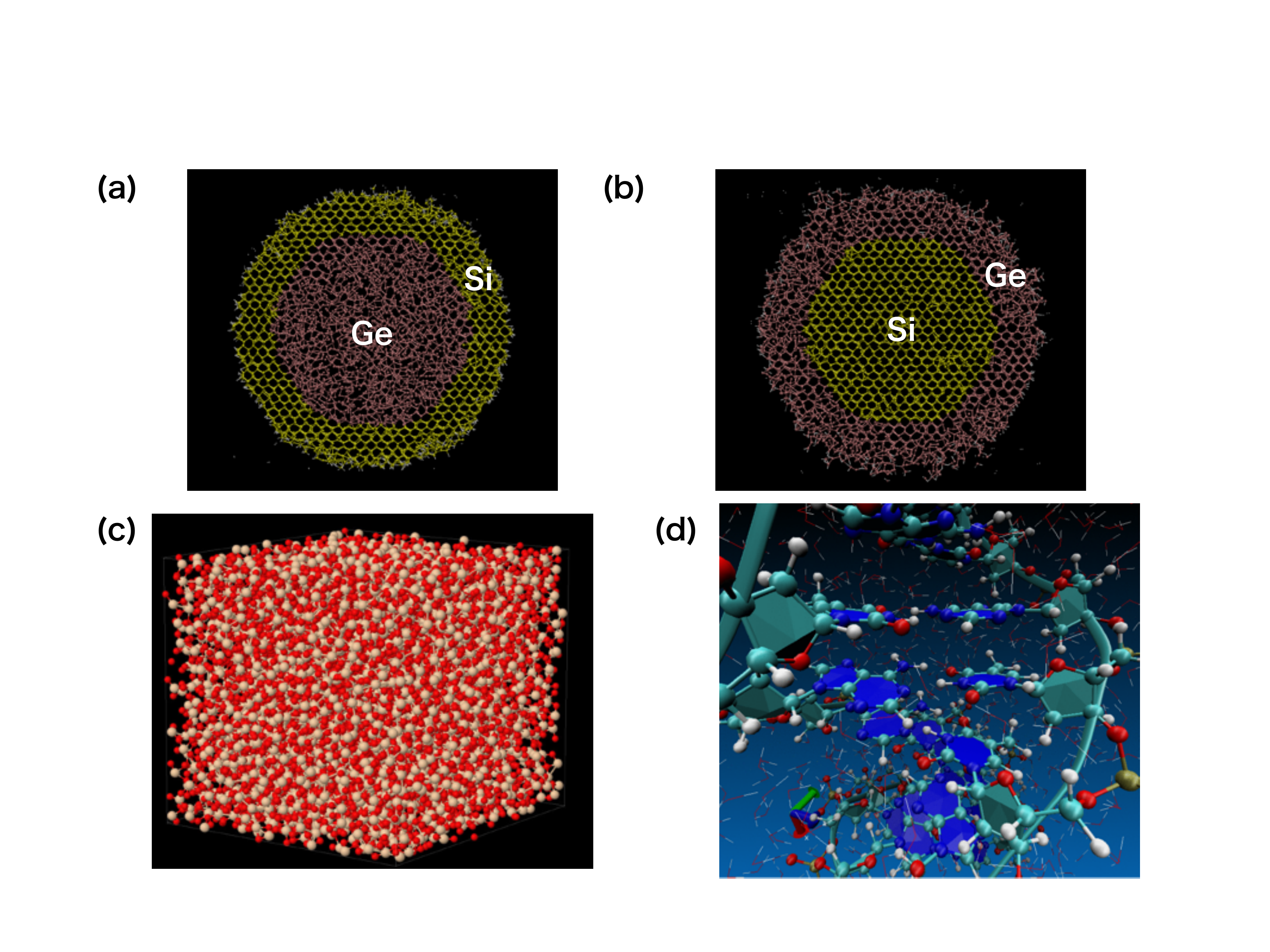}
  \caption{Snapshot structures for (a) Ge/Si core-shell and (b) Si/Ge
    core-shell nanowires at 3000K, (c) melting SiO$_2$ at 3000K and
    10GPa. (d) hydrated DNA system (details of the simulation are
    found in Ref.\protect\onlinecite{Otsuka2016}).\label{fig:MD}}
\end{figure}

For perfect, epitaxial Si/Ge core-shell nanowires,
DFT-MD simulations of nanowires containing 4,788
atoms, whose diameter is 10.4 nm (Si core is 7.2 nm and the thickness
of Ge shell part is 1.6 nm), at 900K were recently performed.  The
DFT-MD simulations confirmed that the structure is stable at least up
to 10 pico seconds.  This does not guarantee that the perfect
epitaxial model is more stable than other structures containing
defects or intermixing, but it indicates that the model is at least a
meta-stable structure.  We also performed DFT-MD simulations of
Si/Ge and Ge/Si core-shell nanowires at 3000K, whose snapshot
structures are shown in Fig.~\protect\ref{fig:MD}(a) and (b).  We
observed that the Ge region melted first in both cases.  As linear-scaling
DFT-MD simulations on such large systems are now practical, we expect
that they can be used to explore possible structures of various types
of defects or intermixing effects at the Si/Ge interfaces by a local
heating technique. Such study is now in progress.  Furthermore, as we
explained in Sec.~\ref{sec:MD}, we can now perform
DFT-MD simulations at a constant high temperature and a given high
pressure.  Structural properties of melting SiO$_2$
(Fig.~\protect\ref{fig:MD}(c)) are now being investigated using
\textsc{Conquest}.

Of course, complex biomolecules, such as DNA in water, are also an
important target for large-scale DFT-MD studies using
\textsc{Conquest}, with a snapshot shown in Fig.~\ref{fig:MD}(d).  It is
noteworthy that free energy calculations based on the blue moon
ensemble method are now available with
\textsc{Conquest}\cite{Hirakawa2020}.  We expect a 
variety of dynamical processes or enzyme reactions in biological
systems will be studied with \textsc{Conquest} in the future.

\section{Conclusions}
\label{sec:conclusions}

We have summarised the principles behind the implementation of the
\textsc{Conquest} code, which enables it to address large scale DFT
simulations, up to around 10,000 atoms with exact diagonalisation,
and significantly larger systems, at least up to millions of atoms
using linear scaling DFT.  We showed how support functions can be
represented in three ways, leading to a powerful approach for
representing the density matrix.  We also gave details on approaches
to find the electronic structure of large systems, even with linear
scaling, and indicated how hybrid DFT methods can be extended to
extremely large systems.

We gave details of atomic movement, particularly molecular dynamics,
and how the implementation and performance is affected by the use of
linear scaling methods.  We demonstrated that accurate, linear scaling
MD is feasible with reasonable computational time, for standard
ensembles (NVE, NVT and NPT, though care is needed with the
calculation of stress and linear scaling).  We then showed the
performance of the MSSF approach, and how it opens up the possibility
of exact diagonalisation simulations with many thousands of atoms.  We also
investigated the parallel performance of the code, both with MSSF and
linear scaling, finding excellent performance including perfect
scaling for certain approaches.  We ended by giving examples of
applications of the code, on systems with sizes ranging from hundreds
of atoms to hundreds of thousands of atoms.

While large scale DFT calculations are challenging, in terms of the
preparation of the system, the computing resources required, and the
analysis of large data sets, it is clear that they are also now feasible
for the majority of users.  It is to be hoped that the size of
most DFT calculations will grow from a few hundred atoms to many
thousands, enabling greater accuracy, and new systems to be addressed.

\begin{acknowledgments}
  The authors would like to express their gratitute to all the
  developers and beta testers of \textsc{Conquest} who have
  contributed to the code over the years.  In particular, Professor
  Mike Gillan initiated and oversaw the development and implementation
  for many years, and his input was invaluable. We also acknowledge,
  in alphabetical order, Michiaki Arita, Veronika Brazdova, Marius
  Buerkle, Rathin Choudhury, Chris Goringe, Eduardo Hernandez, Teruo
  Hirakawa, Chathurangi Kumarasinghe, Conn O'Rourke, Takao Otsuka,
  Alex Sena, Kane Shenton, Teppei Suzuki, Umberto Terranova, Milica
  Todorovic, Lianheng Tong, and Antonio Torralba.

This work was supported by World Premier International Research Centre
Initiative (WPI Initiative) on MaterialsNanoarchitectonics (MANA), 
“Exploratory Challenge on Post-K computer” by MEXT,
and JSPS Grant-in-Aid for ScientificResearch (18H01143, 17H05224, 15H01052).
LT was supported by the BBSRC grant BB/H024217/1, ``Linear Scaling
Density Functional Theory for Biochemistry''. 

Calculations were performed on
the Numerical Materials Simulator at NIMS,
the supercomputer HA8000 system at Kyushu University,
and by using the computational resources of the K computer
provided by the RIKEN Advanced Institute for Computational Science
through the HPCI System Research project (Project ID: hp160129,
hp170264, hp180175, hp180226 and hp190096).

The authors are grateful for computational support from the UK
Materials and Molecular Modelling Hub, which is partially funded by
EPSRC (EP/P020194), for which access was obtained via the UKCP
consortium and funded by EPSRC grant ref EP/P022561/1.  They also
acknowledge computational support from the UK
national high-performance computing service, ARCHER, for which access
was obtained via the UKCP consortium and funded by EPSRC grant ref
EP/K013564/1.

\textbf{Data availability} The data that support the findings of
this study are available from the corresponding authors upon
reasonable request. 

\end{acknowledgments}

\nocite{*}
\bibliography{references}

\appendix

\section{Further details of MD implementation}
\label{sec:further-details-md}

Here, we explain the details of the integration scheme used in the molecular dynamics,
since it is impoartant for the actual implementation, and related to the stability of
the molecular dynaimcs.

\subsection{Microcanonical ensemble\label{sec:micr-ensemble}}

The microcanonical ensemble is generated simply by solving Hamilton's equation
of motion for the Hamiltonian,

\begin{equation}
  \mathcal{H} = \sum_{i=1}^N \frac{\mathbf{p}_i^2}{2m_i} + U(\mathbf{r}_i),
\end{equation}

resulting in the following equations of motion:

\begin{align}
  \mathbf{\dot{r}}_i &= \frac{\mathbf{p}_i}{m_i} \\
  \mathbf{\dot{p}}_i &= \frac{\partial U(\mathbf{r}_i)}{\partial\mathbf{r}_i} = \mathbf{F_i}
\end{align}

These equations are integrated using the velocity Verlet algorithm.

\subsection{Non-Hamiltonian molecular dynamics\label{sec:non-hamilt-molec}}

Hamiltonian dynamics describe systems that are isolated from their surroundings,
and in order to generate the canonical and isobaric-isothermal ensembles, the
system must be coupled to an external bath (heat for theformer and heat and
stress in the case of the latter). In the extended system approach, a set of
\textit{non-Hamiltonian} equations of motion including degrees of freedom for a
thermostat and/or barostat are posited, and shown to generate the correct
statistical ensemble \textit{post hoc}.

\subsubsection{Canonical (NVT) ensemble\label{sec:canon-nvt-ensemble}}

The Nos\'e-Hoover Hamiltonian\citep{Nose1984,Hoover1985} for the canonical
ensemble can be written,

\begin{equation}
  \mathcal{H} = \sum_i \frac{1}{2}m_i s^2\mathbf{\dot{r}}_i^2 + U(\mathbf{r}_i) +
  \frac{1}{2}Q\dot{s}^2 - (n_f + 1)k_B T \ln s,
\end{equation}

where $\mathbf{r}_i$ and $\mathbf{\dot{r}}_i$ are respectively the position and
velocity of particle $i$, $U$ is the potential energy (in this case the DFT
total energy), $s$ is a dimensionless quantity that can be interpreted
\textit{post hoc} as a time step scaling factor, $Q$ is the fictitious mass of
the heat bath and $n_f$ is the number of ionic degrees of freedom. Hamilton's
equations of motion can then be solved to generate the Nos\'e-Hoover equations
of motion. However, Martyna \textit{et al.} demsonstrate that this method does
not generate an ergodic trajectory, and propose an alternative formulation with
a chain of $M$ coupled heat thermostats of mass $Q_k$, each with ``position''
$\eta_k$ and conjugate momentum $p_{\eta_k}$ \citep{Martyna1992}, resulting in
the following equations of motion.

\begin{align}
  \mathbf{\dot{r}_i} &= \frac{\mathbf{p}_i}{m_i} \\
  \mathbf{\dot{p}_i} &= -\frac{\partial U(\mathbf{r})}{\partial \mathbf{r}_i} -
  \frac{p_{\eta_1}}{Q_1}\mathbf{p}_i \\
  \dot{\eta}_k &= \frac{p_{\eta_k}}{Q_k} \\
  \dot{p}_{\eta_1} &= \left(\sum_{i=1}^N\frac{\mathbf{p}_i}{m_i} -
  n_fk_BT\right) -
  \frac{p_{\eta_{2}}}{Q_{\eta_{2}}}p_{\eta_1} \\
  \dot{p}_{\eta_k} &= \left(\frac{p^2_{\eta_{k-1}}}{Q_{k-1}} - k_BT\right) -
  \frac{p_{\eta_{k+1}}}{Q_{k+1}}p_{\eta_k} \\
  \dot{p}_{\eta_M} &= \left(\frac{p^2_{\eta_{M-1}}}{Q_{M-1}} - k_BT\right)
\end{align}

These equations are integrated by constructing an appropriate Liouvillian and
translated into an algorithm via the Trotter-Suzuki expansion, as described in
Hirakawa \textit{et al} \citep{Hirakawa2017}.

\subsubsection{Isobaric-isothermal (NPT) ensemble\label{sec:isob-isoth-npt}}

The Parrinello-Rahman equations of motion \citep{Parrinello1981} extend the
constant volume equations of motion to include the degrees of freedom of the
unit cell via the extended system approach. \textsc{Conquest} uses the
Martyna-Tobias-Tuckerman-Klein modification \citep{Martyna1996}, coupling the
constant pressure equations of motion to a Nos\'e-Hoover chain thermostat to
recover the NPT ensemble. For an cell unconstrained unit cell, the equationf of
motion are,

\begin{align}
  \mathbf{\dot{r}}_i &= \frac{\mathbf{p}_i}{m_i} + \frac{\mathbf{p}_g}{W_g}\mathbf{r}_i \\
  \mathbf{\dot{p}}_i &= \mathbf{F}_i - \frac{\mathbf{p}_g}{W_g}\mathbf{p}_i -
  \left(\frac{1}{N_f}\right)\frac{\mathrm{Tr}[\mathbf{p}_g]}{W_g}\mathbf{p}_i -
  \frac{p_\xi}{Q}\mathbf{p}_i \\
  \mathbf{\dot{h}} &= \frac{\mathbf{p}_g\mathbf{h}}{W_g} \\
  \mathbf{\dot{p}_g} &= V(\mathbf{P}_\mathrm{int}-\mathbf{I}P_\mathrm{ext}) +
  \left[\frac{1}{N_f}\sum_{i=1}^N\frac{\mathbf{p}_i^2}{m_i}\right]\mathbf{I} -
  \frac{p_\xi}{Q}\mathbf{p}_g \\
  \dot{\xi} &= \frac{p_\xi}{Q} \\
  \mathbf{\dot{p}}_g &= \sum_{i=1}^N\frac{\mathbf{p}_i^2}{m_i} +
  \frac{1}{W_g}\mathrm{Tr}[\mathbf{p}_g^T\mathbf{p}_g] - (N_f + d^2)kT,
\end{align}

where $\mathbf{r}_i$, $\mathbf{p}_i$ and $m_i$ are respectively the position, momentum and
mass of particle $i$, $\xi$, $p_\xi$ and $Q$ are the position, momentum and mass
of the thermostat and $\mathbf{h}$, $\mathbf{p_g}$ and $W_g$ are the matrix of
lattice vectors, their velocities and the barostat mass. For simplicity, only a
single Nos\'e-Hoover thermostat is included, but in \textsc{Conquest} a
Nos\'e-Hoover chain is used. The Liouvillian is constructed, and the integrator
constructed using the splitting of Shinoda \textit{et al} \citep{Shinoda2004},

\begin{equation}
  iL = iL_r + iL_h + iL_v + iL_\mathrm{bath},
\end{equation}

which can be further decomposed,

\begin{align}
  iL_\mathrm{bath} &= iL_\mathrm{box} + iL_\mathrm{particles} \\
  iL_\mathrm{box} &= iL_\mathrm{vbox} + iL_\xi + iL_{v_{\xi_1}} + iL_{v_{\xi_k}} + iL_{v_{\xi_M}} \\
  iL_\mathrm{particles} &= iL_\mathrm{vpart} + iL_\xi + iL_{v_{\xi_1}} + iL_{v_{\xi_k}} + iL_{v_{\xi_M}}
\end{align}

Then, using Liouville's theorem, we have,

\begin{align}
  iL_r &= \sum_{i=1}^N[\mathbf{v}_i + \mathbf{v}_g\mathbf{r}_i]\cdot\nabla_{\mathbf{r}_i} \\
  iL_h &= \sum_{\alpha,\beta}\mathbf{v}_{g,\alpha\beta}\mathbf{h}_{\alpha\beta}\frac{\partial}{\partial\mathbf{h}_{\alpha\beta}} \\
  iL_v &= \sum_{i=1}^N\left(\frac{\mathbf{F}_i}{m_i}\right)\cdot\nabla_{\mathbf{v}_i} \\
  iL_\mathrm{bath} &= iL_\mathrm{vpart} + iL_\mathrm{vbox} + iL_\xi +
  iL_{v_{\xi_1}} + iL_{v_{\xi_k}} + iL_{v_{\xi_M}} \\
  &= \sum_{i=1}^N\left[-\left\{\mathbf{v}_g +
    \frac{1}{N_f}\mathrm{Tr}(\mathbf{v}_g) +
    v_{\xi_1}\right\}\mathbf{v}_i\right]\nabla _{\mathbf{v}_i} \nonumber \\
  &+ \sum_{\alpha,\beta}\left[\frac{F_\mathrm{box}}{W} -
    v_{\xi_1}\mathbf{v}_{g,\alpha\beta}\right]\frac{\partial}{\partial\mathbf{v}_{g,\alpha\beta}}
  \nonumber \\
  &+ \sum_{k=1}^M v_{\xi_k}\frac{\partial}{\partial\xi_k} \nonumber \\
  &+ \left[\frac{F_{\mathrm{NHC}_1}}{Q_1} -
    v_{\xi_1}v_{\xi_2}\right]\frac{\partial}{\partial v_{\xi_1}} \nonumber \\
  &+ \sum_{k=2}^M\left[\frac{1}{Q_k}(Q_{k-1}v_{\xi_{k-1}}^2 - kT_\mathrm{ext}) -
    v_{\xi_k}v_{\xi_{k+1}}\right]\frac{\partial}{\partial v_{\xi_k}} \nonumber \\
  &+ \left[\frac{1}{Q_M}(Q_{M-1}v_{\xi_{M-1}}^2 -
    kT_\mathrm{ext})\right]\frac{\partial}{\partial v_{\xi_M}}. \nonumber
\end{align}

In this instance, we use $M$ Nos\'e-Hoover heat baths. The equations of motion
can then be expanded via the Trotter-Suzuki identity, and directly translated
into an algorithm.

\begin{equation}
  e^{iL\Delta t} = e^{iL_\mathrm{bath}\frac{\Delta t}{2}}e^{iL_v\frac{\Delta t}{2}}e^{iL_h\frac{\Delta t}{2}}e^{iL_r\Delta t}e^{iL_h\frac{\Delta t}{2}}e^{iL_v\frac{\Delta t}{2}}e^{iL_\mathrm{bath}\frac{\Delta t}{2}}.
\end{equation}

This integrator is tested on a bulk crystalline silicon system, as explained in Sec.~\ref{sec:MD}.

\subsubsection{Weak-coupling thermostat and barostat\label{sec:weak-coupl-therm}}

The Berendsen weak coupling method \citep{Berendsen1984} involves global
coupling to a pressure and/or heat bath via a Langevin-type equation of motion
with a global friction constant. In the case of the thermostat, the ionic
velocities are rescaled by a factor $\lambda$, which is scaled towards the
target temperature $T_0$ by the coupling frequency $1/\tau_T$.

\begin{equation}
  \lambda = \left[ 1 + \frac{\Delta t}{\tau_T}\left(\frac{T_0}{T}-1\right)\right]^{\frac{1}{2}}
\end{equation}

Similarly, for the barostat, the cell is rescaled by the matrix $\mathbf{\mu}$,
which is scaled towards a target pressure tensor $\mathbf{P}_0$ by the pressure
coupling frequency $1/\tau_P$ and the estimated bulk modulus $\beta$.

\begin{equation}
  \mathbf{\mu} = \mathbf{I} - \frac{\beta\Delta t}{3\tau_P}(\mathbf{P}_0 - \mathbf{P}) 
\end{equation}

While trivial to implement, weak coupling will not generate the correct
canonical or isobaric-isothermal velocity distribution, and the thermostat has
the pathological effect of systematically transferring energy to the most slowly
changing degrees of freedom (the ``flying ice cube'' effect). However, it may be
useful for equilibration.

\subsubsection{Stochastic velocity rescaling\label{sec:stoch-veloc-resc}}

Stochastic velocity rescaling \citep{Bussi2007} is essentially a modification of
the weak coupling method that does not suffer from the flying ice cube effect. A
correctly constructed random force is added to enforce the correct NVT (or NPT)
phase space distribution. The kinetic energy is rescaled such that the change in
kinetic energy between ionic steps is,

\begin{equation}
  dK = (\bar{K} - K)\frac{dt}{\tau} + 2\sqrt{\frac{K\bar{K}}{N_f}}\frac{dW}{\sqrt{\tau}},
\end{equation}

where $\bar{K}$ is the target kinetic energy (i.e. heat bath temperature), $dt$
is the time step, $\tau$ is the time scale of the thermostat, $N_f$ is the
number of degrees of freedom and $dW$ is a Wiener process. In practice, the
particle velocities are resecaled by a factor $\alpha$, defined as,

\begin{eqnarray}
  \alpha^2 &=& e^{-\Delta t/\tau} + \frac{\bar{K}}{N_fK}\left(1-e^{-\Delta
  t/\tau}\right)\left(R_1^2 + \sum_{i=2}^{N_f}R_i^2\right) \nonumber\\
  &+& 2e^{-\Delta
  t/2\tau}\sqrt{\frac{\bar{K}}{N_fK}\left(1-e^{-\Delta t/\tau}\right)R_1},
\end{eqnarray}

where $R_i$ is a set of $N_f$ normally distributed random numbers with unitary
variance. This thermostat can be used in NPT dynamics \citep{Bussi2009} by
barostatting the system via the Parrinello-Rahman method, but with additional
$R_i$'s for the cell degrees of freedom, thermostating the cell velocities as
well as the ionic velocities.

\end{document}